\documentclass[letterpaper,10pt,twocolumn]{article}
\usepackage[T1]{fontenc}
\usepackage[utf8]{inputenc}
\usepackage{authblk}
\usepackage{graphicx}
\usepackage{amsmath}
\usepackage{lscape}
\usepackage{soul}
\usepackage{amssymb}
\usepackage[twocolumn,textwidth=18.00cm,columnsep=.41cm]{geometry}

\usepackage[labelsep=endash]{caption}

\begin{document}

\title{\textbf{Probing exoplanet clouds with optical phase curves}\footnote{Published in PNAS (2015),
vol. 112, no. 44, 13461–13466, doi: 10.1073/pnas.1509135112;
www.pnas.org/cgi/doi/
10.1073/pnas.1509135112.
}}


\author{A. Garc\'ia Mu\~noz ($\ddagger$, \footnote{Currently at Technische Universit\"at Berlin, D-10623 Berlin, Germany.
Contact: tonhingm@gmail.com, garciamunoz@astro.physik.tu-berlin.de.} ) and K. G. Isaak\\
Scientific Support Office, Directorate of Science and Robotic Exploration, ESA/ESTEC, Keplerlaan 1, 2201 AZ, Noordwijk, The Netherlands\\
($\ddagger$) ESA Research Fellow
}

\date{\vspace{-6.5ex}}



\maketitle

\textbf{
Kepler-7b is to date the only exoplanet for which clouds have been
inferred from the optical phase curve—from visible-wavelength
whole-disk brightness measurements as a function of orbital
phase. Added to this, the fact that the phase curve appears dominated
by reflected starlight makes this close-in giant planet a
unique study case. Here we investigate the information on coverage
and optical properties of the planet clouds contained in the
measured phase curve. We generate cloud maps of Kepler-7b and
use a multiple-scattering approach to create synthetic phase curves,
thus connecting postulated clouds with measurements. We show
that optical phase curves can help constrain the composition and size
of the cloud particles. Indeed, model fitting for Kepler-7b requires
poorly absorbing particles that scatter with low-to-moderate anisotropic
efficiency, conclusions consistent with condensates of silicates,
perovskite, and silica of submicron radii. We also show that we are
limited in our ability to pin down the extent and location of the
clouds. These considerations are relevant to the interpretation of
optical phase curves with general circulation models. Finally, we estimate
that the spherical albedo of Kepler-7b over the Kepler passband
is in the range 0.4–0.5.
}


Significance: \textit{
We investigate the potential of optical phase curves for the
characterization of exoplanet clouds. In the case of the low-density,
close-in giant planet Kepler-7b, its phase curve reveals
that the planet clouds are optically thick and composed of poorly
absorbing condensates. We also constrain the cloud particle size,
which is an important parameter in the study of the photochemistry,
wind dynamics, and cloud microphysics of the planet
atmosphere. Our work establishes a valuable framework within
which to investigate exoplanet atmospheres using upcoming flagship
space missions such as the European Space Agency’s CHEOPS
and PLATO and NASA’s TESS, which will all fly within the decade.}


Phase curves provide unique insight into the atmosphere of a planet, a fact well-known and tested in solar system exploration \cite{arkingpotter1968,mallama2009, tomaskosmith1982}.
Disentangling the information encoded in a phase curve is a complex process however, 
and interpretations can be faced with degeneracies. 
The potential of phase curves to characterise exoplanet atmospheres, particularly in combination with other techniques, is tantalising. Phase curves observed over all orbital phases (OPs)
 are available for a few close-in planets in the optical (passband central wavelengths $\lambda$$<$0.8 $\mu$m) 
\cite{angerhausenetal2014,coughlinlopezmorales2012,demoryetal2011,demoryetal2013,demory2014,
estevesetal2013,estevesetal2015, faiglermazeh2011,gelinokane2014,kippingspiegel2011,kippingbakos2011,snellenetal2009} 
and the infrared (1 $\mu$m$\le$$\lambda$$\le$24 $\mu$m)
 \cite{knutsonetal2007, crossfieldetal2010, lewisetal2013, stevensonetal2014}. At infrared wavelengths the measured flux from hot planets is typically dominated by thermal emission. In the optical, both thermal emission and reflected starlight contribute, with the relative size of the contributions 
dependent on the measurement wavelength as well as 
on the temperature of the atmosphere and the occurrence of condensates 
\cite{cahoyetal2010,hengdemory2013,marleyetal1999,schwartzcowan2015,seageretal2000,sudarskyetal2000}. 

Kepler-7b \cite{lathametal2010} is one of the $\sim$1000 planets discovered by the \textit{Kepler} mission. 
Its inferred mass $M_{\rm{p}}$ (=0.44$M_{\rm{J}}$; J for Jupiter) and radius $R_{\rm{p}}$ (=1.61$R_{\rm{J}}$) 
result in an unusually low bulk density (0.14 g cm$^{-3}$) that is inconsistent with current models of giant planet 
interiors \cite{baraffeetal2010,fortneynettelmann2010}. Kepler-7b orbits a quiet G-type star  
of effective temperature $T_{\star}$=5933 K 
every 4.89 days (orbital distance $a$=0.062 astronomical units) \cite{demoryetal2011, demoryetal2013}, and  
tidal forces have likely synchronised its orbit and spin motions. 
Taken together these set a planet equilibrium temperature $T_{\rm{eq}}$$\le$1935 K. 

\textit{Kepler} photometry (0.4--0.9 $\mu$m) of the star-planet system has 
enabled the optical study of Kepler-7b 
\cite{angerhausenetal2014,coughlinlopezmorales2012,demoryetal2011,demoryetal2013, estevesetal2015,kippingbakos2011}. 
The inferred geometric albedo, $A_{\rm{g}}$=0.25--0.38 
\cite{angerhausenetal2014,demoryetal2011,demoryetal2013,estevesetal2015,kippingbakos2011}, 
reveals a planet of reflectivity comparable to the 
solar system giants ($A_{\rm{g}}$=0.4--0.5), which is unexpectedly high for a close-in gas planet. 
Theory indeed predicts that the strong stellar irradiation that a planet in such an orbit experiences 
strips off reflective clouds, rendering the planet dark ($A_{\rm{g}}$$<$0.1) \cite{marleyetal1999,sudarskyetal2000}. 
The prediction is largely consistent with empirical evidence, 
and dark planets dominate the sample of known close-in giant planets
\cite{demory2014,kippingspiegel2011,hengdemory2013,barclayetal2012, roweetal2008}. 
Exceptions exist, and other planets 
(51 Peg b, $A_{\rm{g}}$=0.5$\times$(1.9/($R_{\rm{p}}$/$R_{\rm{J}}$))$^2$ at 0.38--0.69 $\mu$m 
\cite{martinsetal2015}; 
HD 189733b, $A_{\rm{g}}$=0.40$\pm$0.12 at 0.29--0.45 $\mu$m
\cite{evansetal2013}; 
KOI-196b, $A_{\rm{g}}$=0.30$\pm$0.08 at 0.4--0.9 $\mu$m
\cite{santerneetal2011}) with
elevated albedos suggest that we are beginning to sample the diversity of exoplanet atmospheres.
Potentially compensating for strong stellar irradiation, 
Kepler-7b's low surface gravity (417 cm s$^{-2}$) may help sustain 
reflective condensates lofted in the upper atmosphere 
that would increase the planet albedo \cite{sudarskyetal2000}.

Brightness temperatures for Kepler-7b inferred from occultations at 3.6 and 4.5 $\mu$m 
with \textit{Spitzer} ($<$1700 and 1840 K, respectively  \cite{demoryetal2013}) are well below the equivalent
brightness temperature deduced from \textit{Kepler} data ($\sim$2600 K).
This key constraint, placed in the framework of heat recirculation in the atmospheres of close-in giants,
is evidence that the \textit{Kepler} optical phase curve is dominated by reflected starlight 
rather than by thermal emission \cite{demoryetal2013,hengdemory2013,huetal2015}. 
Interestingly, the peak of the optical phase curve occurs after secondary eclipse (OP$>$0.5), 
when the planet as viewed from Earth is not fully illuminated and longitudes westward of the 
substellar point are preferentially probed.  This asymmetry hints at a 
spatial structure in Kepler-7b's envelope caused by horizontally 
inhomogenous clouds \cite{demoryetal2013,hengdemory2013,huetal2015}.  
Subsequent investigations have identified other planets that show similar offset between 
occultation and peak brightness \cite{angerhausenetal2014,estevesetal2015}. 
However the lack of infrared measurements for these means that it has not been
possible to rule out contamination in the optical by a thermal component as the cause of the asymmetry.
 
Recent work has used the optical phase curve of Kepler-7b to build
brightness maps \cite{demoryetal2013,huetal2015}, investigate the
prevalence of reflected starlight over thermal emission \cite{huetal2015},  
and explore plausible cloud configurations \cite{webberetal2015}. 
No previous study has systematically connected the extent, location and optical thickness of the cloud,  
or the composition and size of the suspended particles, to the measured phase curve.
That exercise is the objective of this paper.

\section*{Atmospheric model}

\begin{figure*}[htb]
   \centering
   \includegraphics[width=7.cm]{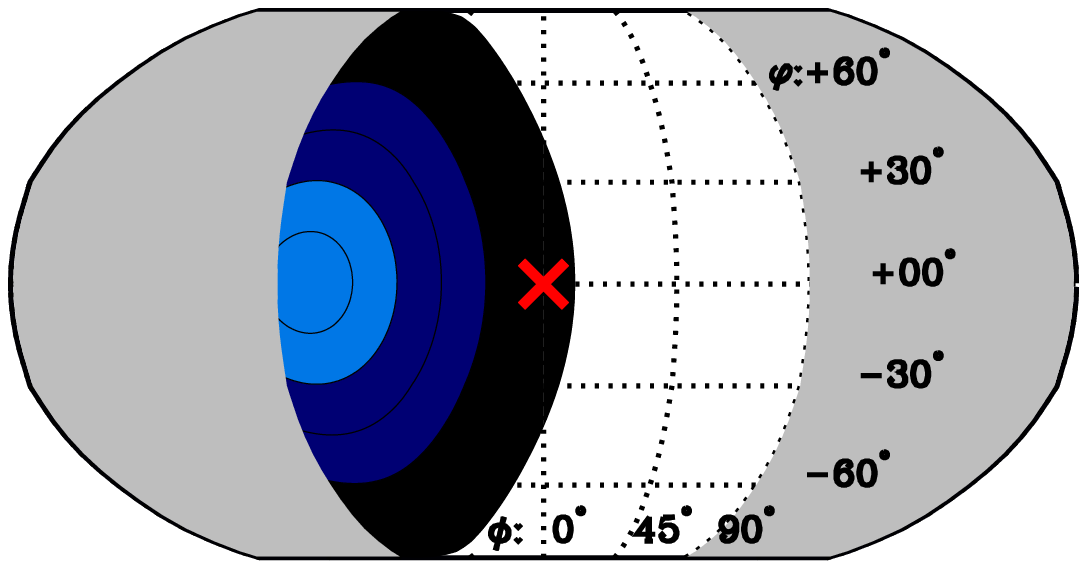}
   \caption{\label{pattern_fig}
   Prescribed cloud map for Kepler-7b.
   In our parameterisation of clouds, 
   the optical thickness drops exponentially 
   from the cloud centre (SI Appendix, Eq. S2);  
   $\sigma_{\rm{c}}$ is a measure of the cloud extent, and 
   $\Delta$$\phi_{\rm{c}}$ is the eastward offset of the cloud 
   relative to the substellar point 
   (denoted by the cross in the centre of the schematic). 
   In the example, 
   $\sigma_{\rm{c}}$=30$^{\circ}$ and
   $\Delta$$\phi_{\rm{c}}$=$-$80$^{\circ}$ (the cloud is west of the substellar
   point). 
   Different colour tones represent various levels of $\tau$/$\tau_{\rm{c}}$. 
   Model input parameters $\varpi_{\rm{0}}$, $g_1$, and $r_{\rm{g}}$ 
   are taken to be uniform throughout the cloud. 
   The cloud is assumed to be fixed with respect to
   the substellar point.
   }
   \end{figure*}

We set up an idealised atmospheric model of Kepler-7b that 
mimics a vertically uniform yet horizontally inhomogeneous cloud atop a gas sphere. 
In an initial exploratory investigation, 
six parameters ($\tau_{\rm{c}}$, $\sigma_{\rm{c}}$, $\Delta$$\phi_{\rm{c}}$, $\varpi_{\rm{0}}$, $g_1$, $r_{\rm{g}}$)  
specify the optical properties of the cloud-gas medium. 
An analytical expression  captures the inhomogeneity of the optical thickness
$\tau$($\phi$, $\varphi$; $\tau_{\rm{c}}$,  $\sigma_{\rm{c}}$, $\Delta$$\phi_{\rm{c}}$)
for the prescribed cloud in the longitude ($\phi$) and latitude ($\varphi$) directions
(SI Appendix). 
Here, $\tau_{\rm{c}}$ denotes the maximum vertically integrated optical thickness within the cloud,
$\sigma_{\rm{c}}$ yields a measure of its horizontal extent, and 
$\Delta$$\phi_{\rm{c}}$ is the cloud offset eastward of the substellar point (Fig. \ref{pattern_fig}). 
The fraction of photons that propagate further after collisions with cloud particles 
is set by the single scattering albedo $\varpi_{\rm{0}}$, whereas 
the photon propagation directions are dictated by a double Henyey-Greenstein (DHG) phase function that has both forward and 
backward lobes \cite{cahoyetal2010,hovenierhage1989}. The shape of the DHG phase function is parameterised by the 
asymmetry parameter $g_1$ of the forward component (SI Appendix). 
A Lambert-like surface of reflectance $r_{\rm{g}}$ at the bottom of the cloud accounts for the effect of scattering from the gas below. 
The values explored for the six parameters are given in Table \ref{library_table}, 
and include $r_{\rm{g}}$=0, 0.1, 0.2 and 0.3 and thus different degrees of absorption by the gas. 
A non-zero $r_{\rm{g}}$ allows for partial back-scattering from the gas, 
and may be a more realistic representation of the atmosphere -- 
alkalis and titanium/vanadium oxides can have broad spectral wings 
and are believed to be the main gas-phase absorbers
over the \textit{Kepler} passband \cite{sudarskyetal2000,lavvasetal2014,spiegeletal2009}. 
The selected $r_{\rm{g}}$ values are motivated by photochemical and cloud formation models over a range of 
temperatures bracketing the conditions on Kepler-7b \cite{marleyetal1999, sudarskyetal2000}.

\section*{Model phase curves}

We created a grid of possible atmospheric configurations that probes $\sim$1.5 million combinations of the six parameters. 
The radiative transfer calculations were performed using a Monte Carlo algorithm specifically designed for 
horizontally inhomogeneous planets \cite{garciamunoz2015,garciamunozmills2015}. 
For each combination we solved the multiple scattering problem and generated a reflected starlight phase curve
as a function of the star-planet-observer phase angle $\alpha$.
Our treatment assumes that the contribution 
to the optical phase curve from thermal emission is minimal, 
which is appropriate for Kepler-7b \cite{demoryetal2013,huetal2015},
and that can be neglected. 
To compare with the measurements, we expressed the model results as
planet-to-star brightness ratios:
\begin{equation}
F_{\rm{p}}/F_{\star}=(R_p/a)^2 A_g \Phi(\alpha), 
\label{fpfstar_eq}
\end{equation} 
with the planet phase function 
normalised such that $\Phi$($\alpha$=0)$\equiv$1. 
$F_{\rm{p}}$ and $F_{\star}$ are the brightness of the planet and star, respectively.
Kepler-7b follows a circular orbit 
of inclination angle $i$=85.2$^{\circ}$ \cite{demoryetal2011}. 
Changes in $\alpha$ through the planet orbit were accounted for through
$\cos{(\alpha)}$=$-\sin{(i)} \cos{(2\pi \rm{OP})}$.

We evaluated the $\chi^2$ statistic, 
weighted by the measurement uncertainties, 
for the difference in  $F_{\rm{p}}/F_{\star}$ between observations and models (SI Appendix). 
By construction, the calculated $\chi^2$ is a six-parameter function 
$\chi^2$($\tau_{\rm{c}}$, $\sigma_{\rm{c}}$, $\Delta$$\phi_{\rm{c}}$, 
 $\varpi_{\rm{0}}$, $g_1$, $r_{\rm{g}}$).   
We excluded from the summation measurements within the transit and secondary eclipse, 
which makes the total number of usable measurements $N$=1,244. 

\section*{Phase curve interpretation}

We found minimum $\chi_{\rm{m}}^2/N$=1.013, 1.009, 1.027 and 1.076 for $r_{\rm{g}}$=0, 0.1, 0.2 and 0.3, respectively, 
with values of the model input and output (albedo) parameters at the minima listed in Table
\ref{summary_table}. 
Individual confidence intervals for each input parameter were estimated from the inequality $\Delta$$\chi^2$= $\chi^2$$-$$\chi_{\rm{m}}^2$$<$15.1  
by optimising all parameters except $r_{\rm{g}}$ and the one being considered  (SI Appendix). Mathematically, the confidence intervals 
bracket the best-matching input parameters with a 99.99\% probability ($\sim$4 SDs) \cite{bevingtonrobinson2003, pressetal1992}. 
For $r_{\rm{g}}$=0, 0.1, 0.2 and 0.3, the condition  $\Delta$$\chi^2$$<$15.1 resulted in a total of 289, 197, 160 and 131 model phase
curves, respectively, which we refer to as the minimal-$\chi^2$ sets. 

\begin{figure*}[h]
   \centering
   \includegraphics[width=18cm]{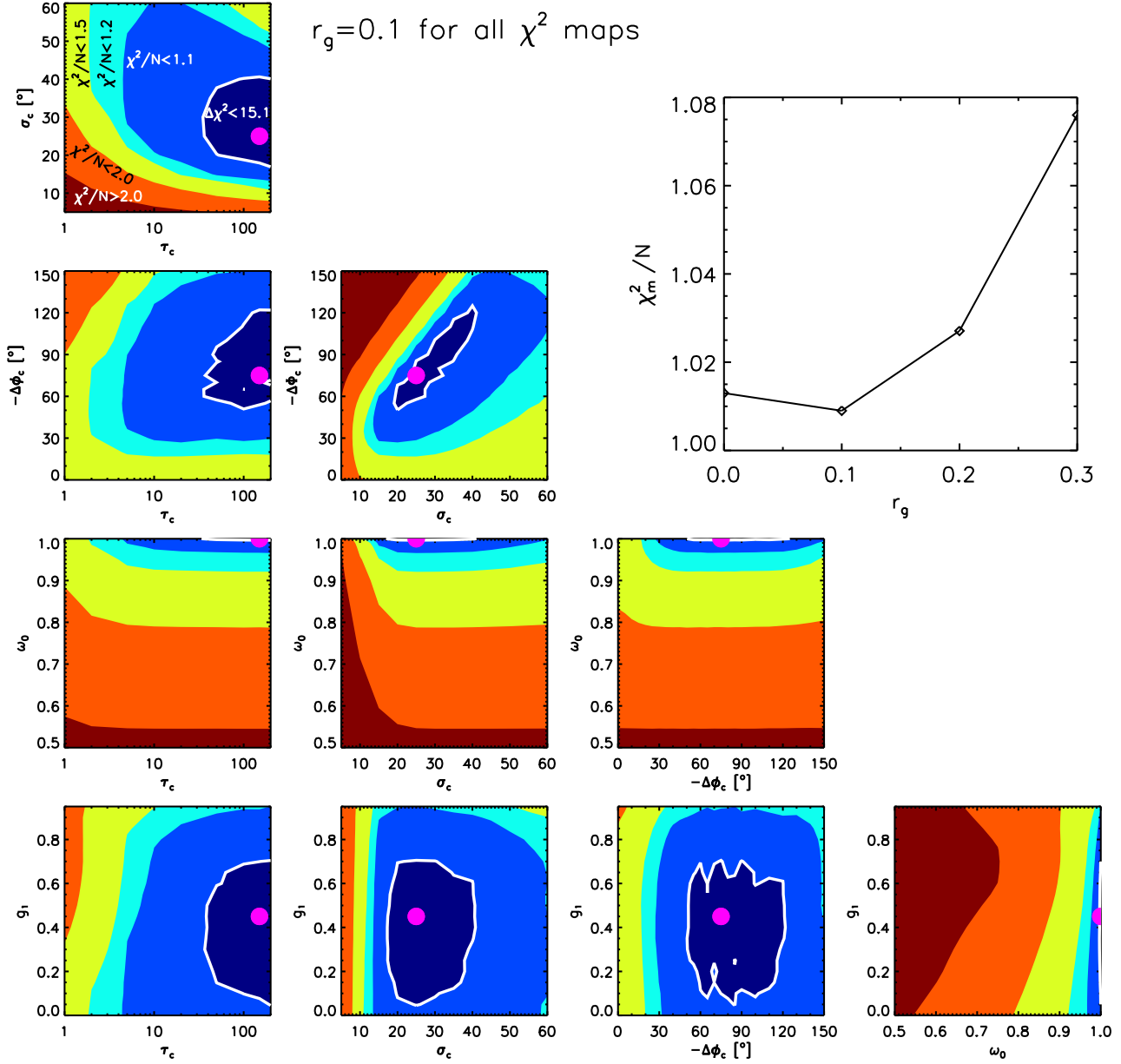}
   \caption{\label{fig2_fig} 
   Projected 2D $\chi^2$ maps for $r_{\rm{g}}$=0.1. Each map is produced by minimising 
   $\chi^2$($\tau_{\rm{c}}$, $\sigma_{\rm{c}}$, $\Delta$$\phi_{\rm{c}}$, $\varpi_{\rm{0}}$, $g_1$, $r_{\rm{g}}$)
   over all parameters but the two being represented and the selected $r_{\rm{g}}$. 
   Contours for $\Delta$$\chi^2$= $\chi^2$$-$$\chi_{\rm{m}}^2$$<$15.1, which defines our minimal-$\chi^2$ sets, are indicated in white.
   Solid colour contours refer to regions with $\chi^2/N$ less/more than the quoted quantities. 
   The magenta circle indicates the corresponding parameter values associated with the best-matching solution and $\chi_{\rm{m}}^2$. 
   Top Right shows variation of $\chi_{\rm{m}}^2$ with surface reflectance $r_{\rm{g}}$. 
   }
   \end{figure*}

Projected 2D $\chi^2$ maps were obtained for each $r_{\rm{g}}$ by optimisation of all input parameters
except the two on the axes and $r_{\rm{g}}$ (Fig. \ref{fig2_fig} for $r_{\rm{g}}$=0.1; SI Appendix, S8--S10 for $r_{\rm{g}}$=0, 0.2 and 0.3).
The maps provide insight into how particular parameter combinations compensate for one another to 
produce degeneracies in the interpretation of the measured phase curve. 
They also help visualise the inferred confidence intervals.

Only prescribed clouds that are optically thick at their centre ($\tau_{\rm{c}}$$\ge$20) 
reproduce the observations well. 
The constant-$\chi^2$($\sigma_{\rm{c}}, \Delta$$\phi_{\rm{c}}$)
contours show that the best-matching 
$\sigma_{\rm{c}}$ and $\Delta$$\phi_{\rm{c}}$ values are correlated in 
continua of ($\sigma_{\rm{c}}$, $\Delta$$\phi_{\rm{c}}$) combinations 
that begin from narrow ($\sigma_{\rm{c}}$$\sim$15$^{\circ}$) clouds
moderately displaced ($\Delta$$\phi_{\rm{c}}$$\sim$$-$45$^{\circ}$; westwards)
from the substellar point and extend to much broader patterns centered 
near or beyond the western terminator. 
This is evidence that a variety of
clouds provide comparable net scattering from the planet's visible day side
and reproduce the brightness peak after secondary eclipse.

We also find that cloud particles 
with near-zero absorption ($\varpi_{\rm{0}}$$\sim$1) 
must be invoked, and that 
low $\chi^2$s are obtained within a broad interval of $g_1$ values. 
The asymmetry parameters for the forward component ($g_1$) and full
scattering phase function ($g$) generally
depend on the shape, size and composition of the scattering particles as well as
on the wavelength of the incident radiation \cite{hansentravis1974}.
The fact that the minimal-$\chi^2$ sets contain model phase curves with
$g_1$ values in the 0.05--0.7 range (0.05--0.57 for $g$; SI Appendix) suggests
that both small particles (leading to isotropic scattering) 
and larger particles (leading to moderate anisotropic scattering efficiency) 
produce model phase curves consistent with the observations. 
Furthermore, the projected $\chi^2$($\Delta$$\phi_{\rm{c}}, g_1$) reveals that 
$\chi^2$ variations are gradual in the $g_1$ direction, 
which makes it difficult to constrain this parameter.
The latter difficulty is inherent to the interpretation of brightness phase curves and is well known from studies of Venus 
\cite{arkingpotter1968}.
The modest signal-to-noise ratio of the Kepler-7b measurements does not create the difficulty, but does exacerbate it (SI Appendix).
These conclusions are generally valid for all $r_{\rm{g}}$=0, 0.1, 0.2 and 0.3
minimal-$\chi^2$ sets.

Interestingly, $\chi_{\rm{m}}^2$($r_{\rm{g}}$=0)$\sim$$\chi_{\rm{m}}^2$($r_{\rm{g}}$=0.1)$<$$\chi_{\rm{m}}^2$($r_{\rm{g}}$=0.2)$<$$\chi_{\rm{m}}^2$($r_{\rm{g}}$=0.3), 
and thus better fits are obtained when the prescribed cloud rests 
above a poorly reflecting gas atmosphere. 
Provided that alkalis dominate the gas-phase absorption, this might be consistent with 
a cloud that lies at pressures less than 10$^{-4}$ bar for which the neutral-to-ion alkali transition likely occurs in the atmospheres of 
close-in giant planets \cite{lavvasetal2014}. 

\begin{figure*}[h]
   \centering
   \includegraphics[width=9cm]{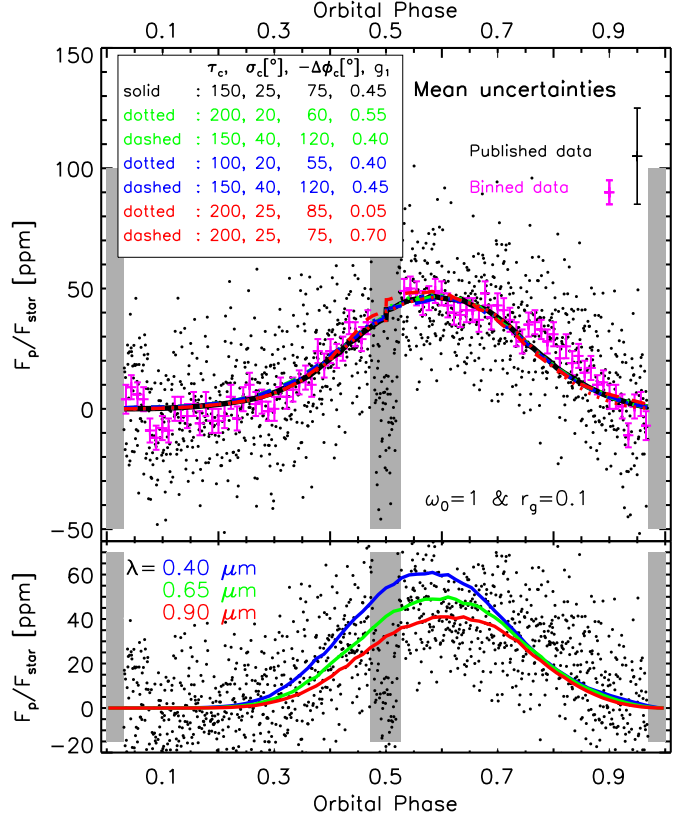}
   \caption{\label{myfits_fig}
   Model phase curves from the minimal-$\chi^2$ sets and predictions for multicolour curves.
   (Top) 
   Observed optical phase curve for Kepler-7b (black dots, measurements, and error bars omitted for clarity;
   magenta, binned data (80 bins) with corresponding error bars) \cite{demoryetal2013}  
   and seven model phase curves (solid lines) within the minimal-$\chi^2$ set
   for $r_{\rm{g}}$=0.1.  
   The selected examples cover a broad range of values for 
   $\sigma_{\rm{c}}$, $\Delta \phi_{\rm{c}}$ and $g_1$.   
   Data points in the grey shaded areas fall within the transit and secondary eclipse and 
   are excluded from the analysis.
   (Bottom) Simulated planet brightness at $\lambda$=0.4, 0.65 and 0.9 $\mu$m based on
   best-matching solutions. Model input parameters are   
  $\sigma_{\rm{c}}$=25$^{\circ}$, $\Delta$$\phi_{\rm{c}}$=$-$65$^{\circ}$, $\varpi_{\rm{0}}$=1, $r_{\rm{g}}$=0, 
  and
  \{$\tau_{\rm{c}}$, $g_1$, $\lambda$\}= 
   \{390, 0.60; 0.4 $\mu$m\},
   \{100, 0.40; 0.65 $\mu$m\}, 
   \{33, 0.20; 0.9 $\mu$m\}.   
   }
   \end{figure*}

The degeneracy in the observation-model fitting is best appreciated in Fig. \ref{myfits_fig},
Top. It confirms that many of the model phase curves in the minimal-$\chi^2$ sets
are virtually indistinguishable to the naked eye,   
even though they correspond to very different cloud configurations.  
In the Venus case, a combination of whole-disk measurements of 
brightness and polarisation 
was required to reveal the composition and size of the upper cloud 
particles \cite{arkingpotter1968,hansenarking1971,hansenhovenier1974}. 

\section*{Cloud particle optical properties}

We attempted to connect the constraints placed 
on the single scattering albedo and asymmetry parameter 
by the minimal-$\chi^2$ sets 
to the composition and particle size of plausible condensates.  
These are critical properties in the study of the photochemistry, 
wind dynamics and cloud microphysics of planetary atmospheres.
We used Mie theory \cite{mishchenkoetal2002} to calculate values for 
$\varpi_{\rm{0}}$ and $g$ using particle parameters such as the effective radius ($r_{\rm{eff}}$) and 
real and imaginary parts of the refractive index 
($n_{\rm{r}}$ and $n_{\rm{i}}$) at a midband wavelength of 0.65 $\mu$m (SI Appendix). 
A key outcome of the Mie calculations is that only condensates with 
$n_{\rm{i}}$$\lesssim$0.003 are consistent with the 
limits of $\varpi_{\rm{0}}$($r_{\rm{eff}}, n_{\rm{r}}, n_{\rm{i}}$)$\gtrsim$0.99 and 
$g$($r_{\rm{eff}}, n_{\rm{r}}, n_{\rm{i}}$)$\lesssim$0.6 that we established 
based on the investigation of the minimal-$\chi^2$ sets.

We identified four candidate cloud components that have both
condensation temperatures of 1,000-2,000 K, relevant to the equilibrium 
temperature of Kepler-7b 
\cite{morleyetal2012,wakefordsing2015}, and $n_{\rm{i}}$$\lesssim$0.003: 
two silicates (Mg$_2$SiO$_4$ and MgSiO$_3$), 
perovskite (CaTiO$_3$) and silica (SiO$_2$). 
Without further information it is not possible to favour one candidate above another.
Due to differences in their refractive indices, 
we were able to tentatively constrain the particle effective radius for each of them.
For the silicates ($n_{\rm{r}}$$\sim$1.6 and $n_{\rm{i}}$$\sim$10$^{-4}$
in the optical,  
www.astro.uni-jena.de/Laboratory/OCDB/), 
we found that only Mie-scattering particles with $r_{\rm{eff}}$=0.04--0.16 $\mu$m complied with the
inferred limits on $\varpi_{\rm{0}}$ and $g$. 
In turn, 
for perovskite ($n_{\rm{r}}$=2.25 and $n_{\rm{i}}$$\sim$10$^{-4}$ \cite{wakefordsing2015,uedaetal1998}),
the range of effective radii is $r_{\rm{eff}}$=0.02--2 $\mu$m. 
And for silica ($n_{\rm{r}}$=1.5 and $n_{\rm{i}}$$\sim$10$^{-7}$ \cite{kitamuraetal2007}),
$r_{\rm{eff}}$=0.004--0.16 $\mu$m. 

Significant optical thicknesses and
moderately small particle sizes suggest that Kepler-7b's cloud 
might be vertically extended and 
reach high in the atmosphere \cite{webberetal2015}.
This idea appears consistent with the expectations for low-gravity planets \cite{sudarskyetal2000}
and with a plausible cloud base at the 10$^{-4}$ bar level. 
Vigorous atmospheric winds, as predicted by 3D models \cite{parmentieretal2013}, 
could well keep micron-sized and smaller particles aloft at such pressures \cite{hengdemory2013}.

It is appropriate to verify the above conclusions with a full Mie multiple-scattering treatment. 
The DHG approximation is a convenient but simplified approach to particle scattering. 
A more thorough treatment of the multiple scattering problem would have used particle 
phase functions derived from, e.g., Mie theory  
and would have explored separately the impact of $n_{\rm{r}}$, $n_{\rm{i}}$ and $r_{\rm{eff}}$ 
on the model phase curves. 
At present, such a treatment is computationally prohibitive, which justifies our 
adoption of the DHG parameterisation for the exploration exercise. 

To assess the conclusions on the inferred values of $r_{\rm{eff}}$, 
we produced model phase curves specific to silicate, perovskite, and silica clouds. 
The new grid  differs from the one summarised in Table \ref{library_table} mainly in that 
we now omit the DHG parameterisation. Instead, we implement
particle scattering phase functions and single scattering albedos $\varpi_{\rm{0}}$
obtained directly from Mie theory. 
For each particle composition (and corresponding $n_{\rm{r}}$, $n_{\rm{i}}$ values), 
we sample $r_{\rm{eff}}$ from 0.001 to 100 $\mu$m to produce the needed input to the
multiple scattering problem. By forming
$\chi^2$($\tau_{\rm{c}}$, $\sigma_{\rm{c}}$, $\Delta$$\phi_{\rm{c}}$,  
$r_{\rm{eff}}$, $r_{\rm{g}}$) for each composition, and confidence intervals
$\Delta$$\chi^2$$<$15.1 for each model input parameter, 
we confirm that the above conclusions based on the DHG parameterisation 
are overall valid. 
Specifically, for $r_{\rm{eff}}$ we infer confidence intervals of 0.1--0.32 $\mu$m (silicates), 
0.08--0.2 $\mu$m (perovskite) and 0.1--0.4 $\mu$m (silica). 
Smaller particles result in excessive absorption, whilst particles that are too large lead to unobserved back scattering
at small phase angles in the planet phase curves (SI Appendix, Fig. S3).
These particle radii are to be preferred over those found 
from the DHG parameterisation.

\section*{Planet albedos}

Both the geometric $A_{\rm{g}}$ and spherical $A_{\rm{s}}$ albedos 
quantify the overall reflecting properties of a planet (SI Appendix). 
For the minimal-$\chi^2$ sets, 
we infer $A_{\rm{g}}$$\sim$0.2--0.3 and $A_{\rm{s}}$$\sim$0.4--0.5 
(Table \ref{summary_table}).
An $A_{\rm{s}}$$\sim$0.5 means that 
Kepler-7b reflects about half of the visible-wavelength radiation that it
receives.
Most of the stellar output for $T_{\star}$=5933 K is emitted over the \textit{Kepler} passband, 
and thus the derived spherical albedos are also first-order approximations to the 
Bond albedos $A_{\rm{B}}$ that impact directly on the energy budget 
and equilibrium temperature of the planet. 
Taking $A_{\rm{B}}$$\sim$0.5, we obtain 
an estimate for the planet equilibrium temperature (assuming no heat recirculation) 
of $T_{\rm{eq}}$=1935/2$^{1/4}$$\sim$1630 K. 

\section*{Multi-colour phase curves}

Our investigation of Kepler-7b has shown that high-precision broadband optical photometry can help characterise 
exoplanet atmospheres.
In the coming decade, photometric missions  
such as \textit{CHEOPS} \cite{fortieretal2014}, 
\textit{PLATO} \cite{raueretal2014} and 
\textit{TESS} \cite{rickeretal2015}
will provide optical phase curves of numerous targets. 
Over the same period, 
the \textit{James Webb Space Telescope} will provide the necessary
infrared-discrimination power to separate the planet components due to reflected starlight and thermal emission.
Future exoplanet missions will then be able to obtain multi-colour and spectrally-resolved phase curves.
 
We have explored the added value of multi-colour data. 
We produced model phase curves at $\lambda$=0.4 $\mu$m, 0.65 $\mu$m, and 0.9 $\mu$m 
based on our best-matching solutions of the Kepler-7b data. 
Again, we assumed that thermal emission at all wavelengths considered is negligible 
and accounted for wavelength-dependent changes in $\tau_{\rm{c}}$ and $g_1$ by assuming that 
the dominating cloud particles have $r_{\rm{eff}}$=0.1 $\mu$m, $n_{\rm{r}}$=1.5 and $n_{\rm{i}}$=0.
The synthetic phase curves (Fig. \ref{myfits_fig}, Bottom) and the
differences between them of up to 20 ppm suggest that multi-colour observations  
will provide additional constraints on cloud properties.

\section*{Relevance to General Circulation Models}

There are other approaches to interpret optical phase curves. 
One of these is based on general circulation models (GCMs), 
which jointly treat the dynamics, energetics and chemistry of 3D atmospheres
\cite{knutsonetal2007,hengetal2011,mayneetal2014,showmanetal2015}. 
GCMs often omit clouds however, a simplification 
likely to affect the simulated atmospheric fields. 
In addition, omitting clouds will affect the overall planet 
brightness, as clouds 
potentially reflect much of the visible-wavelength incident starlight. 
Our finding that a continuum of cloud patterns is consistent with Kepler-7b's optical phase curve suggests that
more than one GCM solution will also reproduce a given observed optical phase curve.
We have shown that difficult-to-constrain factors such as the extent and location of the cloud pattern 
(dependent on the condensation of available atmospheric substances  
and on whether the cloud forms locally or is transported from the night side by planet-scale winds)
or the asymmetry parameter (dependent on the cloud particle microphysics)
may compensate for one another and result in visually equivalent optical phase curves.
The credibility of GCM-based predictions of optical phase curves is 
sensitive to these considerations: ignoring them
may lead to the misinterpretation of available and future optical data.
\\

\textbf{Acknowledgments}

We acknowledge Brice-Olivier Demory (University of
Cambridge, Cambridge, UK) for kindly providing the Kepler-7b data, 
Agust\'in S\'anchez-Lavega and Santiago P\'erez-Hoyos (Universidad del Pa\'is Vasco/Euskal Herriko Unibertsitatea (UPV/EHU), Spain) for ideas on confidence
intervals, and Christian Schneider (European Space Agency/European Space
Research and Technology Centre (ESA/ESTEC), the Netherlands) for a critical
reading of the manuscript and discussions on the estimation of uncertainties.
A.G.M. gratefully acknowledges generous ESA support through an ESA
Research Fellowship.



\cleardoublepage


\cleardoublepage

\begin{table*}[ht]
\caption{
\label{library_table}
Input parameter values in the atmospheric model for the exploratory investigation 
(DHG scattering phase function within the cloud).
Our grid of synthetic phase curves contains 1,512,000 combinations of the six parameters.
}            
\centering   
\begin{tabular}{ c  c  c }
Parameter 	& Grid values & \multicolumn{1}{c}{Total}	 \\ 
\hline
$\tau_{\rm{c}}$ 					&  1, 2, 5, 10, 20, 50, 100, 150, 200 & \multicolumn{1}{c}{9} \\
$\sigma_{\rm{c}}$ [$^{\circ}$]  		&  5$\rightarrow$60, step of 5 	& \multicolumn{1}{c}{12}	    \\
$-\Delta$$\phi_{\rm{c}}$ [$^{\circ}$] 	& 0$\rightarrow$90, step of 5;100$\rightarrow$150, step of 10 	 & \multicolumn{1}{c}{25}               \\
$\varpi_{\rm{0}}$ 				&  0.50, 0.75, 0.90, 0.95, 0.98, 0.99, 1 	& \multicolumn{1}{c}{7} \\
$g_1$ 						&  0$\rightarrow$0.95, step of 0.05 & \multicolumn{1}{c}{20} \\
$r_{\rm{g}}$ &  0, 0.1, 0.2, 0.3				& \multicolumn{1}{c}{4}			                          \\
\hline
\multicolumn{2}{}{} &\multicolumn{1}{c}{1,512,000} \\\cline{3-3}
\end{tabular}
\end{table*}

\begin{table*}[h]
\caption{\label{summary_table}
Model input/output parameters for selected phase curves.  
}   
\centering                          
\begin{tabular}{ c  c  c c  c  c c  c  c }
\cline{2-9} 
\multicolumn{1}{c}{} & \multicolumn{5}{c}{Model input parameters} & \multicolumn{3}{c}{Inferred output parameters}\\
\hline
Comment & $\tau_{\rm{c}}$ & $\sigma_{\rm{c}}$ & $-$$\Delta$$\phi_{\rm{c}}$ & $\varpi_{\rm{0}}$ & $g_1$ &  $\chi^2/N$  & $A_{\rm{g}}$ & \multicolumn{1}{c}{$A_{\rm{s}}$} \\
        &                 &           [$^{\circ}$]   &       [$^{\circ}$] &                   &       &             &                 & \multicolumn{1}{c}{}               \\
\hline \hline
Best; $r_{\rm{g}}$=0.0 & 100 & 25 & 65 & 1 & 0.40 & 1.013 & 0.274 & \multicolumn{1}{c}{0.455} \\
\hline
Minimal-$\chi^2$ set; $r_{\rm{g}}$=0.0 & 20--200 & 20--45 & 45--120 & 1 & 0.1--0.7 & $\le$1.025 & 0.24--0.31 & \multicolumn{1}{c}{0.42--0.49} \\
\hline\hline
Best; $r_{\rm{g}}$=0.1                           & 150 & 25 & 75 & 1 & 0.45 &  1.009 & 0.275 & \multicolumn{1}{c}{0.461} \\
\hline
Minimal-$\chi^2$ set; $r_{\rm{g}}$=0.1 & 50--200 & 20--40 & 55--120 & 1 & 0.05--0.7 & $\le$1.021 & 0.24--0.30 & \multicolumn{1}{c}{0.43--0.49} \\
\hline\hline
Best; $r_{\rm{g}}$=0.2                           & 150 & 25 & 80 & 1 & 0.50 &  1.027 & 0.276 & \multicolumn{1}{c}{0.487} \\
\hline
Minimal-$\chi^2$ set; $r_{\rm{g}}$=0.2 & 50--200 & 20--35 & 60--120 & 1 & 0.1--0.65 & $\le$1.039 & 0.25--0.31 & \multicolumn{1}{c}{0.46--0.51} \\
\hline\hline
Best; $r_{\rm{g}}$=0.3                           & 200 & 25 & 90 & 1 & 0.50 &  1.076 & 0.295 & \multicolumn{1}{c}{0.512} \\
\hline
Minimal-$\chi^2$ set; $r_{\rm{g}}$=0.3 & 50--200 & 15--35 & 60--120 & 1 & 0.15--0.6 & $\le$1.088 & 0.28--0.32 & \multicolumn{1}{c}{0.49--0.54} \\
\hline
\end{tabular}
\end{table*}


\begin{thebibliography}{100}

\bibitem[1]{arkingpotter1968} 
Arking, A. \& Potter, J. (1968)
The phase curve of Venus and the nature of its clouds.
J. Atmos. Sci., 25:617--628.

\bibitem[2]{mallama2009} 
Mallama, A. (2009)
Characterization of terrestrial exoplanets based on the phase curves and albedos of Mercury, Venus and Mars.
Icarus, 204:11--14.

\bibitem[3]{tomaskosmith1982} 
Tomasko, M.G. \&
Smith, P.H. (1982)
Photometry and polarimetry of Titan: Pioneer 11 observations and their implications for aerosol properties.
Icarus, 51:65--95.

\bibitem[4]{angerhausenetal2014} 
Angerhausen, D.,
DeLarme, E. \&
Morse, J. A.
A comprehensive study of Kepler phase curves and secondary eclipes - Temperatures and albedos of confirmed Kepler giant planets.
Publ. Astron. Soc. Pac, \textit{in press}.

\bibitem[5]{coughlinlopezmorales2012} 
Coughlin, J.L. \&
L\'opez-Morales, M. (2012)
A uniform search for secondary eclipses of hot Jupiter in Kepler Q2 light
curves.
Astrophys. J., 143:id.39.

\bibitem[6]{demoryetal2011} 
Demory, B.-O.,
Seager, S.,
Madhusudhan, N.,
Kjeldsen, H., 
Christensen-Dalsgaard, J., et al. (2011)
The high albedo of the hot Jupiter Kepler-7 b.
Astrophys. J. Lett., 735:L12.

\bibitem[7]{demoryetal2013} 
Demory, B.-O.,
de Wit, J.,
Lewis, N.,
Fortney, J.,
Zsom, A., et al. (2013)
Inference of inhomogeneous clouds in an exoplanet atmosphere.
Astrophys. J. Lett., 776:L25.

\bibitem[8]{demory2014} 
Demory, B.-O. (2014)
The albedos of Kepler's close-in super-Earths.
Astrophys. J. Lett., 789:L20.

\bibitem[9]{estevesetal2013} 
Esteves, L.J.,
De Mooij, E.J.W. \&
Jayawardhana, R. (2013)
Optical phase curves of Kepler exoplanets.
Astrophys. J., 772:id.51.

\bibitem[10]{estevesetal2015} 
Esteves, L.J.,
De Mooij, E.J.W. \&
Jayawardhana, R. (2015).
Changing phases of alien worlds: Probing atmospheres of Kepler planets with high-precision photometry.
Astrophys. J., 804:id.150. 

\bibitem[11]{faiglermazeh2011} 
Faigler, S. \&
Mazeh, T. (2011)
Photometric detection of non-transiting short-period low-mass companions through the beaming, ellipsoidal and reflection effects in Kepler and CoRoT light curves.
Mon. Not. R. Astron. Soc., 415:3921--3928.

\bibitem[12]{gelinokane2014}
Gelino, D.M. \&
Kane, S.R. (2014)
Phase curves of the Kepler-11 multi-planet system.
Astrophys. J., 787:id.105.

\bibitem[13]{kippingspiegel2011} 
Kipping, D.M. \&
Spiegel, D.S. (2011)
Detection of visible light from the darkest world.
Mon. Not. R. Astron. Soc., 417:L88.

\bibitem[14]{kippingbakos2011} 
Kipping, D. \& Bakos, G. (2011)
An independent analysis of Kepler-4b through Kepler-8b.
Astrophys. J. Lett., 730:id.50.

\bibitem[15]{snellenetal2009} 
Snellen, I.A.G.,
de Mooij, E.J.W. \&
Albrecht, S. (2009)
The changing phases of extrasolar planet CoRoT-1b.
Nature, 459:543--545.

\bibitem[16]{knutsonetal2007} 
Knutson, H.A.,
Charbonneau, D.,
Allen, L.E.,
Forntey, J.J.,
Agol, E. et al. (2007)
A map of the day-night contrast of the extrasolar planet HD 189733b.
Nature, 447:183--186.

\bibitem[17]{crossfieldetal2010} 
Crossfield, I.J.M.,
Hansen, B.M.S.,
Harrington, J.,
Cho, J. Y.-K., 
Deming, D. et al. (2010) 
A new 24 $\mu$m phase curve for $v$ Andromedae b.
Astrophys. J., 723:1436--1446.

\bibitem[18]{lewisetal2013} 
Lewis, N.K.,
Knutson, H.A.,
Showman, A.P.,
Cowan, N.B.,
Laughlin, G. et al. (2013) 
Orbital phase variations of the eccentric giant planet HAT-P-2b.
Astrophys. J., 766:id.95.

\bibitem[19]{stevensonetal2014} 
Stevenson, K.B.,
D\'esert, J.-M.,
Line, M.R.,
Bean, J.L.,
Fortney, J.J. et al. (2014)
Thermal structure of an exoplanet atmosphere from phase-resolved emission spectroscopy.
Science, 346:838--841.

\bibitem[20]{cahoyetal2010} 
Cahoy, K.L.,
Marley, M.S. \&
Fortney, J.J. (2010)
Exoplanet albedo spectra and colors as a function of planet Phase, separation, and metallicity.
Astrophys. J., 724:189--214.

\bibitem[21]{hengdemory2013} 
Heng, K. \&
Demory, B.-O. (2013)
Understanding trends associated with clouds in irradiated exoplanets.
Astrophys. J., 777:id.100.

\bibitem[22]{marleyetal1999} 
Marley, M.S.,
Gelino, C.,
Stephens, D.,
Lunine, J. \&
Freedman, R. (1999)
Reflected spectra and albedos of extrasolar giant planets. I. Clear and cloudy atmospheres.
Astrophys. J., 513:879--893.

\bibitem[23]{schwartzcowan2015} 
Schwartz, J.C. \&
Cowan, N.B.
Balancing the energy budget of short-period giant planets:
Evidence for reflective cloud and optical absorbers. 
MNRAS, 449:4192-4203.

\bibitem[24]{seageretal2000} 
Seager, S.,
Whitney, B.A. \&
Sasselov, D.D. (2000)
Photometric light curves and polarization of close-in extrasolar giant planets.
Astrophys. J., 540:504--520.

\bibitem[25]{sudarskyetal2000} 
Sudarsky, D.,
Burrows, A. \&
Pinto, P. (2000)
Albedo and reflection spectra of extrasolar planets.
Astrophys. J., 538:885--903.

\bibitem[26]{lathametal2010} 
Latham, D.W.,
Borucki, W.J.,
Koch, D.G.,
Brown, T.M., 
Buchhave, L.A. et al. (2010)
Kepler-7b: A transiting planet with unusually low density.
Astrophys. J. Lett., 713:L140.

\bibitem[27]{baraffeetal2010} 
Baraffe, I.,
Chabrier, G. \&
Barman, T. (2010)
The physical properties of extra-solar planets.
Reports on Progress in Physics, 73:id.016901.

\bibitem[28]{fortneynettelmann2010} 
Fortney, J.J. \&
Nettelmann, N. (2010)
The interior structure, composition, and evolution of giant planets.
Space Science Reviews, 152:423--447.

\bibitem[29]{barclayetal2012} 
Barclay, T.,
Huber, D.,
Rowe, J.F.,
Fortney, J.J.,
Morley, C.V. et al. (2012)
Photometrically derived masses and radii of the planet and star in the TrES-2
system.
Astrophys. J., 761:id.53.

\bibitem[30]{roweetal2008} 
Rowe, J.F.,
Matthews, J.M.,
Seager, S.,
Miller-Ricci, E.,
Sasselov, D. et al. (2008)
The very low albedo of an extrasolar planet: MOST space-based photometry of HD 209458.
Astrophys. J., 689:1345--1353.

\bibitem[31]{martinsetal2015} 
Martins, J.H.C., Santos, N.C., Figueira, P., et al. (2015)
Evidence for spectroscopic direct detection of reflected light from 51 Peg b?
Astron. \& Astrophys., 576:A134.

\bibitem[32]{evansetal2013} 
Evans, T.M, 
Pont, F.,
Sing, D.K.,
Aigrain, S.,
Barstow, J.K., et al. (2013)
The Deep Blue Color of HD 189733b: Albedo Measurements with Hubble Space Telescope/Space Telescope Imaging Spectrograph at Visible Wavelengths.
Astrophys. J. Lett., 772:L16.

\bibitem[33]{santerneetal2011} 
Santerne, A., 
Bonomo, A.S., 
H\'ebrard, G., 
Deleuil, M., 
Moutou, C., et al. (2011)
SOPHIE velocimetry of Kepler transit candidates. 
IV. KOI-196b: a non-inflated hot Jupiter with a high albedo.
Astron. \& Astrophys., 
536:id.A70.

\bibitem[34]{huetal2015} 
Hu, R.,
Demory, B.-O.,
Seager, S.,
Lewis, N. \& Showman, A. (2015)
A semi-analytical model of visible-wavelength phase curves of exoplanets and applications to Kepler-7 b and Kepler-10 b.
Astrophys. J., 802:51.

\bibitem[35]{webberetal2015} 
Webber, M.W., 
Lewis, N.K.,
Marley, M.,
Morley, C.,
Fortney, J. \&
Cahoy, K. (2015)
Effect of Longitude-dependent Cloud Coverage on Exoplanet Visible Wavelength Reflected-light Phase Curves.
Astrophys. J., 804:id.94. 

\bibitem[36]{hovenierhage1989} 
Hovenier, J.W. \&
Hage, J.I. (1989)
Relations involving the spherical albedo and other photometric quantities of planets with thick atmospheres.
Astron. \& Astrophys., 214:391--401.

\bibitem[37]{lavvasetal2014} 
Lavvas, P.,
Koskinen, T. \&
Yelle, R.V. (2014)
Electron densities and alkali atoms in exoplanet atmospheres.
Astrophys. J., 796:id.15.

\bibitem[38]{spiegeletal2009} 
Spiegel, D.S.,
Silverio, K. \&
Burrows, A. (2009)
Can TiO explain thermal inversions in the upper atmospheres of irradiated giant planets?
Astrophys. J., 699:1487--1500.

\bibitem[39]{garciamunoz2015} 
Garc\'ia Mu\~noz, A. (2015)
Towards a comprehensive model of Earth's disk-integrated Stokes vector. 
Int. J. Astrobiology, 14:379--390.

\bibitem[40]{garciamunozmills2015} 
Garc\'ia Mu\~noz, A. \&
Mills, F.P. (2015)
Pre-conditioned backward Monte Carlo solutions to radiative 
transport in planetary atmospheres. 
Fundamentals: Sampling of propagation directions in polarising media.
Astron. \& Astrophys., 
573:A72.

\bibitem[41]{bevingtonrobinson2003}
Bevington, P.R. \&
Robinson, D.K. (2003)
Data reduction and error analysis for the physical sciences.
McGraw Hill.

\bibitem[42]{pressetal1992}
Press, W.H.,
Teukolsky, S.A.,
Vetterling, W.T. \&
Flannery, B.P. (2003)
Numerical recipes in Fortran 77.
Cambridge University Press.

\bibitem[43]{hansentravis1974} 
Hansen, J.E. \&
Travis, L.D. (1974)
Light scattering in planetary atmospheres.
Space Science Reviews, 16:527--610.

\bibitem[44]{hansenarking1971} 
Hansen, J.E. \&
Arking, A. (1971)
Clouds of Venus: Evidence for their nature.
Science, 171:669--672.

\bibitem[45]{hansenhovenier1974} 
Hansen, J.E. \&
Hovenier, J.W. (1974)
Interpretation of the polarization of Venus.
J. Atmos. Science, 31:1137--1160.

\bibitem[46]{mishchenkoetal2002}
Mishchenko, M.I.,
Travis, L.D. \&
Lacis, A.A. (2002)
Scattering, absorption and emission of light by small particles.
Cambridge Univ. Press.

\bibitem[47]{morleyetal2012} 
Morley, C.V., 
Fortney, J.J., 
Marley, M.S., 
Visscher, C., 
Saumon, D. \& 
Leggett, S.K. (2012)
Neglected Clouds in T and Y Dwarf Atmospheres.
Astrophys. J., 756:id.172.

\bibitem[48]{wakefordsing2015} 
Wakeford, H.R. \& Sing, D.K. (2015)
Transmission spectral properties of clouds for hot Jupiter exoplanets.
Astron. \& Astrophys., 573:id.A122.

\bibitem[49]{uedaetal1998}
Ueda, K., 
Yanagi, H., 
Noshiro, R., 
Hosono, H. \&
Kawazoe, H. (1998)
Vacuum ultraviolet reflectance and electron energy loss spectra of CaTiO$_3$.
J. Phys.: Condens. Matter, 10:3669--3677.

\bibitem[50]{kitamuraetal2007} 
Kitamura, R., 
Pilon, L. \& 
Jonasz, M. (2007)
Optical constants of silica glass from extreme ultraviolet to far infrared at near room temperature.
Appl. Optics, 46:8118--8133.

\bibitem[51]{parmentieretal2013} 
Parmentier, V.,
Showman, A.P. \&
Lian, Y. (2013)
3D mixing in hot Jupiters atmospheres
I. Application to the day/night cold trap in HD 209458b.
Astron. \& Astrophys., 558:id.A91.

\bibitem[52]{fortieretal2014} 
Fortier, A.,
Beck, T.,
Benz, W.,
Broeg, C.,
Cessa, V.
et al. (2014)
CHEOPS: A space telescope for ultra-high precision photometry of exoplanet transits.
Proc. SPIE 9143, Space Telescopes and Instrumentation, 91432J.

\bibitem[53]{raueretal2014} 
Rauer, H., 
Catala, C., 
Aerts, C., 
Appourchaux, T., 
Benz, W.
et al. (2014) 
The PLATO 2.0 mission. 
Experimental Astronomy, 38:249-330.

\bibitem[54]{rickeretal2015} 
Ricker, G.R., 
Winn, J.N., 
Vanderspek, R., 
Latham, D.W., 
Bakos, G.A.
et al. (2015)
Transiting Exoplanet Survey Satellite (TESS). 
J. Astronomical Telescopes, Instruments, and Systems, 1:id.014003.

\bibitem[55]{hengetal2011} 
Heng, K.,
Menou, K. \&
Phillipps, P.J. (2011)
Atmospheric circulation of tidally locked exoplanets: a suite of benchmark tests for dynamical solvers.
Mon. Not. R. Astron. Soc., 413:2380--2402.

\bibitem[56]{mayneetal2014} 
Mayne, N.J., 
Baraffe, I.,
Acreman, D.M.,
Smith, C.
Browning, M.K. et al. (2014)
The unified model, a fully-compressible, non-hydrostatic, deep atmosphere global circulation model, applied to hot Jupiters. ENDGame for a HD 209458b test case.
Astron. \& Astrophys., 561:id.A1.

\bibitem[57]{showmanetal2015} 
Showman, A.P.,
Lewis, N.K. \&
Fortney, J.J. (2015)
3D Atmospheric Circulation of Warm and Hot Jupiters.
Astrophys. J., 801:id.95.







\end{thebibliography}
\end{document}


\title{\textbf{Probing exoplanet clouds with optical phase curves.\\
 Supporting Information Appendix.}}

\author[1,2]{A. Garc\'ia Mu\~noz}
\affil[1]{Scientific Support Office, Directorate of Science and Robotic Exploration, European Space Research and Technology Centre (ESA/ESTEC), Keplerlaan 1, 2201, AZ Noordwijk, The Netherlands}
\affil[2]{ESA Research Fellow}

\author[1]{K. G. Isaak}

\date{\vspace{-5ex}}

\maketitle

\newpage
\textbf{Equilibrium temperature}

The equilibrium temperature of a planet is defined by the equation \cite{trauboppenheimer2010}:
\begin{equation}
T_{\rm{eq}}=\left( \frac{1- A_{\rm{B}}}{4 f} \right)^{1/4} \left( \frac{R_{\star}}{a}
\right)^{1/2} T_{\star}.
\label{teq_eq}
\end{equation}
An upper limit $T_{\rm{eq}}$$\le$1935 K for Kepler-7b is obtained by assuming a Bond
albedo $A_{\rm{B}}$=0 and a heat redistribution factor 
$f$=1/2 (appropriate to a tidally locked planet with no heat circulation),
in addition to $R_{\star}$=2.02$R_{\odot}$ and $T_{\star}$= 5933 K
for the stellar radius and temperature respectively, 
and $a$=0.062 AU for the orbital distance \cite{demoryetal2011}.  
A planet with a non-zero Bond albedo has an equilibrium
temperature $(1-A_{\rm{B}})^{1/4}$ below that of the $A_{\rm{B}}$=0 limit.
\\

\textbf{Atmospheric model}

We assumed  the functional form:  
\begin{equation}
\tau(\phi, \varphi; \tau_{\rm{c}}, \sigma_{\rm{c}}, \Delta\phi_{\rm{c}})= 
\tau_{\rm{c}} \exp[-(\{\phi-\Delta \phi_{\rm{c}}\}^2 + \varphi^2)/(2\sigma^2_{\rm{c}})]. 
\label{tau_eq}
\end{equation}
for the cloud optical thickness (Fig. 1). 
Longitude ($\phi$) and latitude ($\varphi$) are measured from the 
substellar point, which is the relevant coordinate reference for tidally-locked planets. 
The prescribed two-dimensional Gaussian shape 
allows us to readily investigate 
the parameter space of maximum optical thickness, $\tau_{\rm{c}}$, 
cloud width, $\sigma_{\rm{c}}$, and 
cloud offset from the substellar point, $\Delta$$\phi_{\rm{c}}$ ($>$0 eastward). 
This idealised representation of exoplanet clouds contains 
a minimum of essential parameters that may affect the planet appearance in
reflected starlight. 

In our initial exploratory investigation, 
scattering by cloud particles is accounted for by a double Henyey-Greenstein (DHG) phase function \cite{cahoyetal2010,hovenierhage1989}:
\begin{equation}
p_{\rm{DHG}}(\theta; f_1, g_1, g_2)=f_1 p_{\rm{SHG}}(\theta; g_1) + (1-f_1) p_{\rm{SHG}}(\theta; g_2),
\label{dhg_eq}
\end{equation}
where $\theta$ denotes the scattering angle (and $\theta$=0 refers to forward propagation), 
and the single Henyey-Greenstein phase function is given by:
\begin{equation}
p_{\rm{SHG}}(\theta; g)= \frac{1-g^2}{[1+g^2-2g\cos{\theta}]^{3/2}}.
\end{equation}
The DHG phase function is a simple yet valid approximation to more elaborate and physically-founded formulations of particle scattering.
In particular, $p_{\rm{DHG}}$ includes both forward and backward scattering lobes \cite{hovenierhage1989,dlugachyanovitskij1974}
which is critical when modelling the planet phase curve over the entire range of orbital phases.
In order to reduce the number of parameters in $p_{\rm{DHG}}$ and in turn in the atmospheric model of Kepler-7b, 
it was further assumed that $g_2$=$-g_1/2$ and $f_1$=$1-g^2_2$ \cite{cahoyetal2010}. 
This \textit{ad hoc} simplication effectively splits $p_{\rm{DHG}}$ into a 
mainly forward component of asymmetry parameter $g_1$
and a mainly backward component of asymmetry parameter $g_2$, 
and assigns a weight to each of them. 
This behaviour qualitatively resembles more realistic formulations 
such as Mie theory \cite{hansentravis1974}.
Following simple manipulations, 
the asymmetry parameter for the DHG phase function is derived as:
\begin{equation}
g=\mbox{<}\cos{\theta}\mbox{>}=\frac{1}{2} \int_{-1}^{+1} p_{\rm{DHG}}(\theta) \cos{\theta} d(\cos{\theta})= g_1
(1-\frac{3}{8}g^2_1),
\label{gDHG_eq}
\end{equation} 
which relates $g$ to the asymmetry parameter of the forward component 
(and input parameter for the atmospheric model) $g_1$. 
For $g_1$$\in$[0, 1], one has that $g$$\in$[0, 0.625].\\ 

\begin{figure}[h]
\centering
\includegraphics[width=12.5cm]{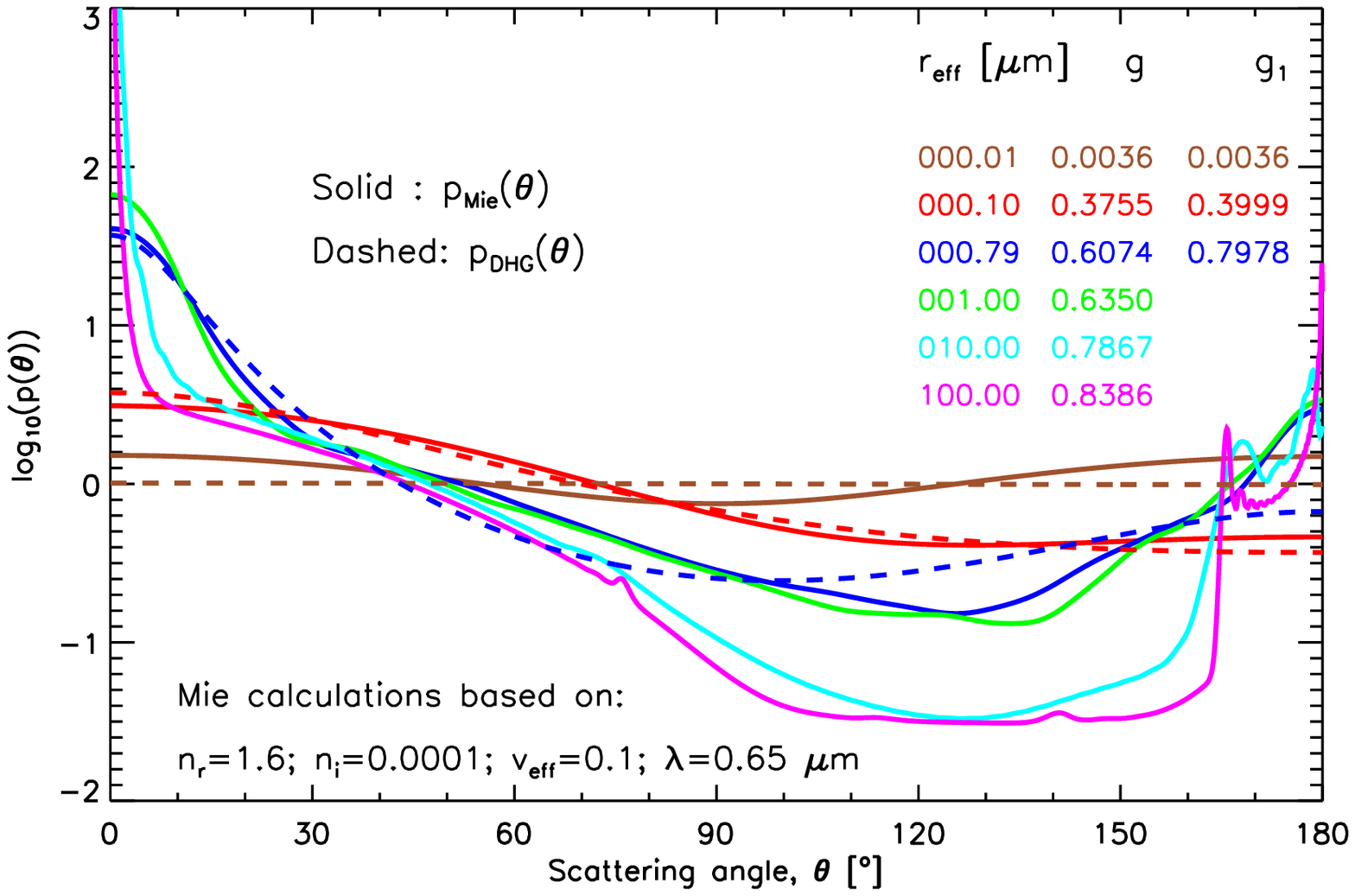}
\caption*{\label{dhgmie_fig} 
Figure S1. 
\textit{Particle scattering phase functions.}
Examples of both $p_{\rm{DHG}}$ and $p_{\rm{Mie}}$ for diverse values of the asymmetry parameter $g$. 
Tabulated: For each $g$ (middle column), the corresponding $r_{\rm{eff}}$ (Mie) and $g_1$ (DHG). 
}
\end{figure}

Alternatively, 
in our full Mie treatment of the multiple-scattering problem, 
particle scattering is accounted for by phase functions obtained from Mie theory, 
$p_{\rm{Mie}}$. 
This more elaborate treatment 
is used to verify the conclusions obtained with the DHG parameterisation in the specific cases of 
silicate, perovskite and silica condensates. 
Figure S1 shows a few $p_{\rm{DHG}}$ and $p_{\rm{Mie}}$ functions for equal values of the asymmetry parameter $g$.
For given values of $n_{\rm{r}}$, $n_{\rm{i}}$ and $v_{\rm{eff}}$, $p_{\rm{Mie}}$ and its corresponding
$g$ depend only on $r_{\rm{eff}}$.
\\


\textbf{Radiative transfer modelling}

We calculated $A_{\rm{g}} \Phi$($\alpha$) 
for each combination of the six  parameters in the atmospheric model of Kepler-7b
(five parameters in the full Mie treatment of multiple scattering)
with a Pre-conditioned Backward Monte Carlo (PBMC) 
algorithm that solves the radiative transfer equation for multiple scattering \cite{garciamunoz2015, garciamunozmills2015, garciamunozetal2014}. 
The PBMC algorithm directly 
considers the integration over the visible disk of the planet, 
which saves computational time when only the disk-integrated signal
is required.
The calculations were carried out in uniform 
steps of 5$^{\circ}$ for the star-planet-observer phase angle $\alpha$, resulting in 
a total of 73 points to sample the entire [-180$^{\circ}$,+180$^{\circ}$]
range of phase angles.

The accuracy of the algorithm has been thoroughly tested with more than 32,000 test cases in previous work 
\cite{garciamunozmills2015} in both its scalar and vector (polarising) formulations,
and for viewing geometries that either spatially resolve the planet or that consider the whole-disk signal. 
As an additional verification, in the course of the current work
we also compared the output of the algorithm to that for optically thick atmospheres 
that scatter according to both (scalar) Rayleigh and SGH phase functions. We found excellent 
agreement for $A_{\rm{g}}$ with published solutions \cite{dlugachyanovitskij1974} (their Table XXI), 
except in one specific case (SHG; $g$=0.85; $\varpi_{\rm{0}}$=0.999). In that instance, 
our solution matched exactly the solution quoted in a later work \cite{hovenierhage1989}, 
which also reported disagreement with Ref. \cite{dlugachyanovitskij1974}.
In addition, we verified that in the limit of zero optical thickness
the numerical solutions 
tended exactly to the analytical expression for Lambert spheres. 
As a final check, 
we compared the model phase curves obtained with the PBMC algorithm to 
the phase curves produced with a classical Gaussian-Tschebychev
disk-integration scheme \cite{horak1950} coupled to a 
plane-parallel radiative transfer solver \cite{stamnesetal1988}. 
In the comparison, we focused on about 400 atmospheric configurations 
specific to horizontally inhomogeneous clouds; particle scattering was
described by either the DHG parameterisation or by the 
full Mie treatment. The agreement was excellent in all cases, 
which covered very diverse cloud-atmosphere conditions. 
The PBMC algorithm clearly outpaces the classical approach when 
the particle scattering phase function becomes moderately asymmetric 
and a large number of streams must be utilised in the classical approach. 

All computations were carried out on `The Grid', the $\sim$80-core computer cluster at ESA/ESTEC--SSO.
As with all Monte Carlo algorithms, the computational cost varies strongly with the number of scattering collisions that
photons undergo before they are either absorbed or escape the medium.
This translates into an increasing computational burden as the medium becomes optically thick and conservative. 
For planets with uniform gas (Rayleigh) atmospheres, Fig. 11 of Ref. \cite{garciamunozmills2015} reports computational times 
for a range of conditions (optical thickness, single scattering albedo, surface reflectance). 
Computational times for more elaborate atmospheric configurations (including Mie scattering and patchy clouds) are reported 
in Table 1 of Ref. \cite{garciamunoz2015}. 
The times for a specified phase angle range from a fraction of a second to a few seconds for 
optical thicknesses of up to $\sim$10. Configurations with non-uniform cloud coverage
run faster than the associated configurations with uniform optical thickness $\tau_{\rm{c}}$ all over the planet.

The majority of multiple scattering calculations with the PBMC algorithm
were conducted with 10,000 photon simulations per phase angle, which results in
accuracies typically better than a few percent. 
The convergence rate of the PBMC algorithm is sensitive to the 
asymmetry of the scattering phase function \cite{garciamunoz2015}, however, and to ensure convergence 
in all cases, we utilised 100,000 photons per phase angle for $g_1$$\ge$0.9.

Figures S2 (DHG) and S3 (Mie) demonstrate with a few examples the sensitivity of the synthetic phase curves to the
model input parameters. 

\begin{figure}[h]
\centering
\includegraphics[width=8.5cm]{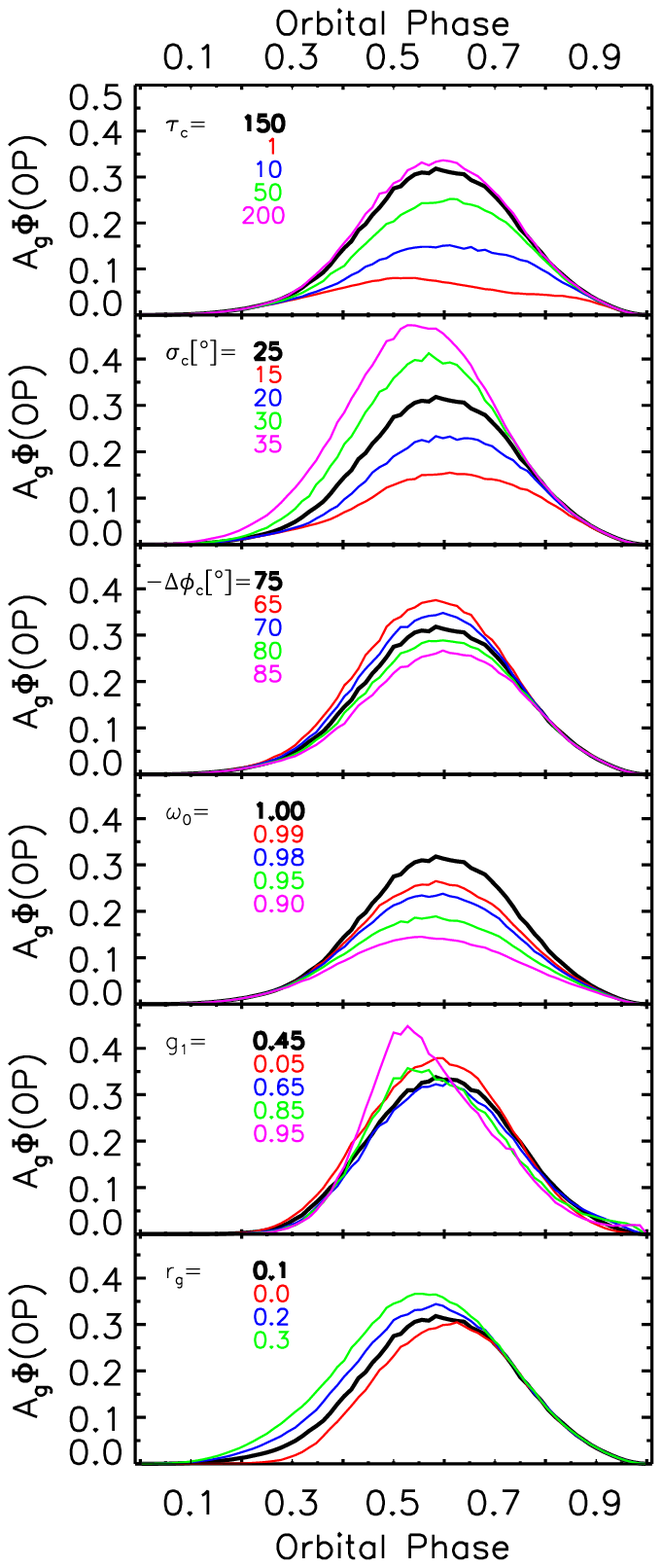}
\caption*{\label{sensi_fig} 
Figure S2. 
\textit{Sensitivity of synthetic phase curves to model input parameters. DHG parameterisation.}
Starting from our best-matching solution (Table 1; black colour, 
bold typeface in the legends), each set of curves shows the result of 
perturbing a single input parameter whilst keeping the others unchanged.
}
\end{figure}

\begin{figure}[t]
\centering
\includegraphics[width=8.5cm]{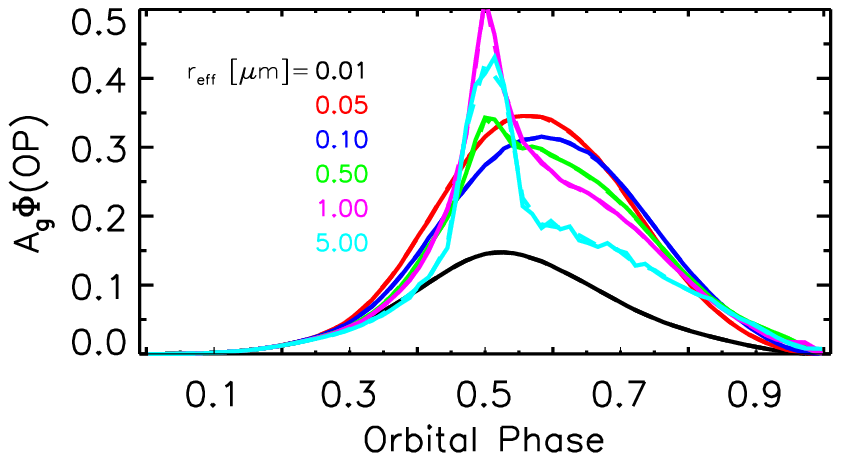}
\caption*{\label{sensimie_fig} 
Figure S3. 
\textit{Sensitivity of synthetic phase curves to the particle effective radius. Full Mie treatment for multiple scattering.} 
The calculations were conducted with 
$\tau_{\rm{c}}$=150, 
$\sigma_{\rm{c}}$=25$^{\circ}$, 
$\Delta$$\phi_{\rm{c}}$=$-$75$^{\circ}$,
$r_{\rm{g}}$=0.1, optical properties of the cloud particles based on 
silicate condensates ($n_{\rm{r}}$=1.6, $n_{\rm{i}}$=10$^{-4}$), and $v_{\rm{eff}}$=0.1
for the particle size distribution. 
Changing $r_{\rm{eff}}$ affects both the particle scattering phase function (Fig. S1) and the 
single scattering albedo (Fig. S6). 
Each curve in the graph actually comprises two overlapping curves, one produced with the PBMC algorithm \cite{garciamunozmills2015} and another one
produced with a Gaussian-Tschebychev disk-integration scheme (16 points in each direction) \cite{horak1950} coupled to a plane-parallel radiative
transfer solver \cite{stamnesetal1988}. 
The difference between the two approaches is typically less than the line thickness.
}
\end{figure}

For a given model phase curve, we infer the geometric albedo $A_{\rm{g}}$ by specifiying 
$\alpha$=0 in Eq. 1 (main text). The spherical albedo 
$A_{\rm{s}}$ is defined as the fraction of monochromatic incident starlight that is
scattered over all directions, and is obtained through numerical integration:
\begin{equation}
A_{\rm{s}}=\int_{-\pi}^{+\pi} A_{\rm{g}} \Phi(\alpha) \sin{(|\alpha|)} d\alpha,
\label{sphericalalbedo_eq}
\end{equation}
over the full range of the star-planet-observer phase angle $\alpha$.
In the case of a Lambert sphere, the ratio $A_{\rm{s}}$/$A_{\rm{g}}$=3/2. 
However, the ratio takes different values 
for non-Lambert scattering (see specific cases in Table 2).

Kepler-7b is assumed to follow a circular orbit (eccentricity$\sim$0.001 \cite{demoryetal2011}) 
of inclination angle $i$=85.2$^{\circ}$ (Fig. S4 top). 
The phase angle and orbital phase are related through
the expression $\cos{(\alpha)}$=$-\sin{(i)} \cos{(2\pi \rm{OP})}$. OP ranges from 0 to 1, while  
$\alpha$ ranges from +180$^{\circ}$ to -180$^{\circ}$ (Fig. S4 bottom).
\\

\begin{figure}[b]
\centering
\includegraphics[width=7.cm]{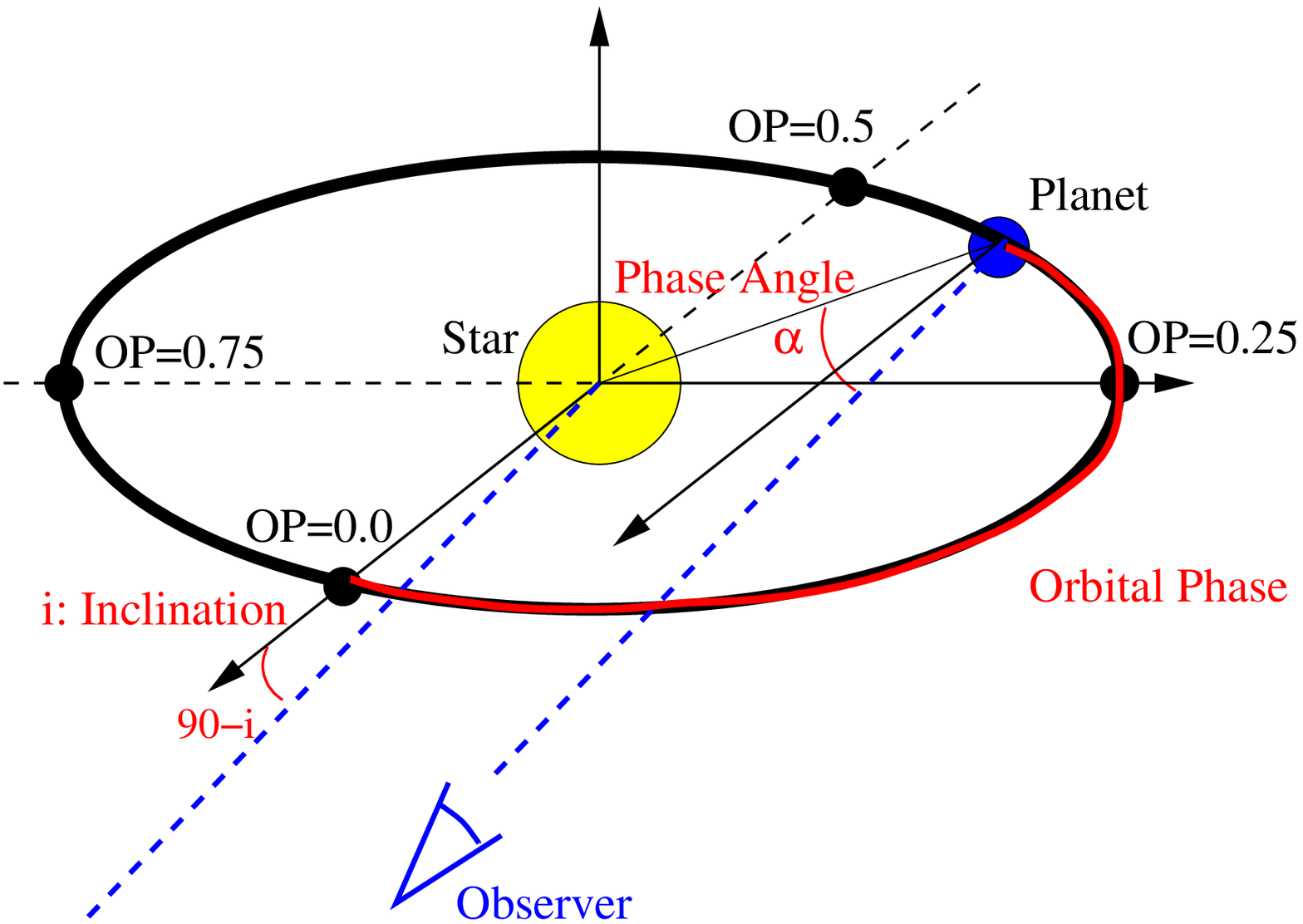}
\includegraphics[width=9.cm]{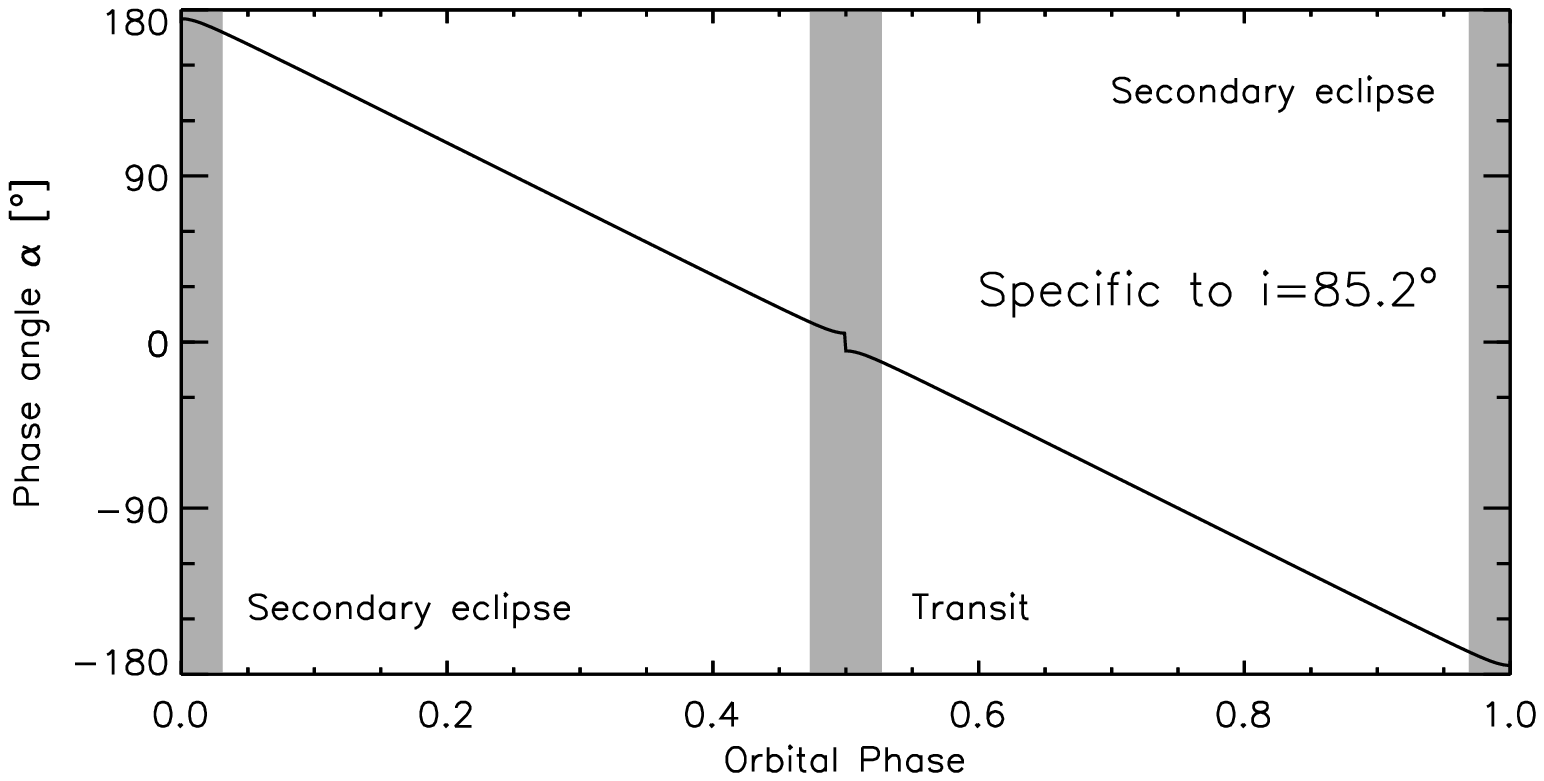}
\caption*{\label{orbitsketch_fig} 
Figure S4. 
\textit{Viewing and orbital geometry.}
Top. 
Schematic showing the definitions of the star-planet-observer phase angle $\alpha$, orbital phase OP and inclination angle $i$. In our implementation:
$\alpha$>0 for OP<0.5, and $\alpha$<0 for OP>0.5. 
Bottom. Evolution of phase angle $\alpha$ with orbital phase OP. Data points within the transit and secondary eclipse (grey shaded areas)
are excluded from the analysis. For $i$$\neq$90$^{\circ}$, $\alpha$ does not go through zero.
}
\end{figure}

\textbf{Goodness of fit. Confidence intervals for the model parameters.}

We utilise $\chi^2$=$\sum_{i}^{N}
[(F_{\rm{p}}/F_{\star}|_{\rm{obs},i}-F_{\rm{p}}/F_{\star}|_{\rm{mod},i})/\sigma_{F_{\rm{p}}/F_{\star}}|_{\rm{obs},i}]^2$ 
as our goodness-of-fit estimator, 
where script $_{\rm{i}}$ refers to individual measurements for the planet-to-star brightness ratio and
$\sigma_{F_{\rm{p}}/F_{\star}}|_{\rm{obs},i}$ are the associated experimental uncertainties. 
We considered only OPs$\in$[0.031, 0.473] and $\in$[0.527, 0.969] in the evaluation of $\chi^2$, 
excluding measurements within the transit and secondary eclipse. 
For each of the $N$=1,244 remaining measurement points, we obtained 
$F_{\rm{p}}/F_{\star}|_{\rm{mod},i}$ by interpolating the model phase curves at the OP of the measurement.

We built projected two-dimensional $\chi^2$ maps for each of the $r_{\rm{g}}$=0,
0.1, 0.2 and 0.3 values in our grid. They are presented in Fig. 2 of the main text
for $r_{\rm{g}}$=0.1, and in Figs. S8-S10 here for $r_{\rm{g}}$=0, 0.2 and 0.3. 
For each pair of parameters and selected $r_{\rm{g}}$, we minimised  
$\chi^2$($\tau_{\rm{c}}$, $\sigma_{\rm{c}}$, $\Delta$$\phi_{\rm{c}}$, $\varpi_{\rm{0}}$, $g_1$, $r_{\rm{g}}$)   
over the remaining three dimensions. 
We estimated confidence intervals for each individual input parameter  
by further projecting on each dimension the iso-contours of the two-dimensional maps 
and finding the boundaries for which the one-dimensional projection satisfies $\Delta$$\chi^2$$<$15.1 
with respect to the minimum, $\chi_{\rm{m}}^2$.
Defined in this manner, the confidence intervals bracket the optimal combination of 
model input parameters that minimises $\chi^2$ with a 99.99\% probability \cite{bevingtonrobinson2003, pressetal1992}. 
In the $\tau_{\rm{c}}$ direction the constant-$\chi^2$ contours
do not form closed patterns within the range of parameter values that we explored, 
which means that clouds with a larger optical thickness $\tau_{\rm{c}}$ at their centre, 
but shifted farther westward, might provide comparably low $\chi^2$s. 
This has no impact on the conclusions on the most relevant parameters for 
characterisation of the cloud and underlying gas, namely: $\varpi_{\rm{0}}$, $g_1$, $r_{\rm{g}}$.

Shown in Fig. S5 are stick plots 
that give the frequency of each model input parameter (top) and estimated albedos (bottom)
within the minimal-$\chi^2$ sets.

   \begin{figure*}
   \centering
   \includegraphics[width=9.cm]{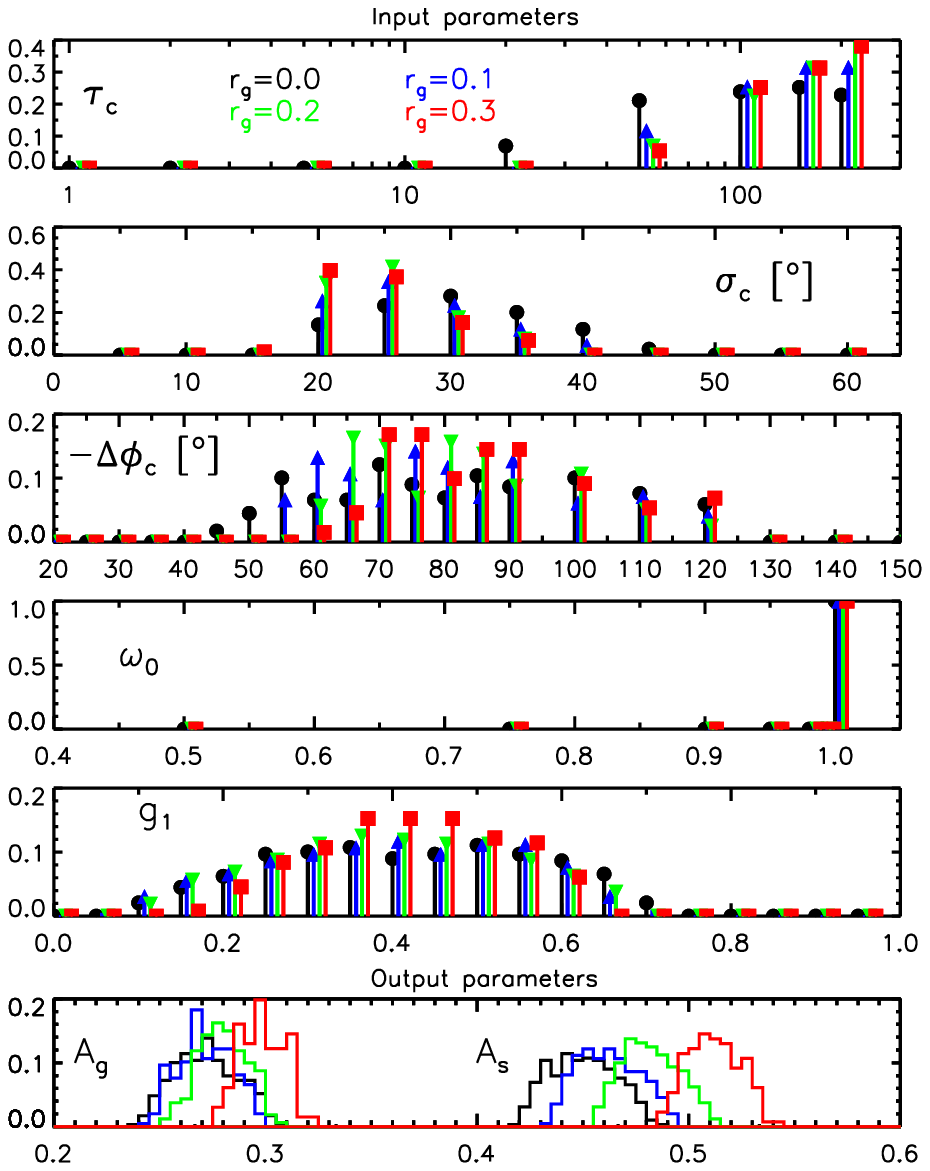}
   \caption*{
   Figure S5. \textit{Investigation of the minimal-$\chi^2$ sets in Table 2.}
   Top.
   Frequencies for input parameters of the atmospheric model for 
   $r_{\rm{g}}$=0 (black, $\chi^2/N$$\le$1.025), 
              0.1 (blue,  $\chi^2/N$$\le$1.021), 
              0.2 (green, $\chi^2/N$$\le$1.039), 
          and 0.3 (red,   $\chi^2/N$$\le$1.088). 
   Sticks for $r_{\rm{g}}$$>$0 are slightly shifted to facilitate their visualisation. 
   Bottom. Occurrence
   frequencies of albedos $A_{\rm{g}}$ (left) and $A_{\rm{s}}$ (right). 
   The frequencies
   are normalised to the number of elements in each
   minimal-$\chi^2$ set (289, 197, 160 and 131, respectively).  
   }
   \end{figure*}

\cleardoublepage

We conducted numerical experiments to assess the impact of measurement 
uncertainties on the retrieval of atmospheric properties and in turn on the degeneracy of the
inversion problem.  
We used as a starting point our best-matching solution for $r_{\rm{g}}$=0 sampled at the OPs of 
the observations (Table 2).
We then added Gaussian noise at the one-sigma level of 5, 10 and 20 ppms for each of three experiments
(for comparison the Kepler-7b measurements have a median uncertainty of 19 ppm), 
and formed the corresponding minimal-$\chi^2$ sets by comparing the so-produced experimental phase curves
against the full grid of model phase curves. 
We then tried to recover the inserted synthetic phase curve, finding
that a broad range of parameter combinations
provided $\chi^2$ deviations less than two percent relative to the corresponding minima.
In the case of noise at the 5 ppm level this meant retrieved parameters: 
$\tau_{\rm{c}}$=50--200; $\sigma_{\rm{c}}$=20--30$^{\circ}$; $-\Delta$$\phi_{\rm{c}}$=50--90$^{\circ}$; 
$\varpi_{\rm{0}}$=1; $g_1$=0.1--0.65. 
The range of model parameters that reproduce the simulated experiments
is even broader for noise at the 10 and 20 ppm level. 
The numerical experiments reveal that small measurement uncertainties trigger 
the intrinsic degeneracy 
associated with the interpretation of brightness phase curves.
\\



\textbf{Mie theory calculations}

We investigated the impact of composition and size of the cloud particles 
on both their single scattering albedo ($\varpi_{\rm{0}}$) and asymmetry parameter ($g$)
with a Mie theory code \cite{hansentravis1974}. 
We assumed for the particle sizes a power law distribution of effective variance $v_{\rm{eff}}$=0.1, 
which removes much of the structure 
associated with mono-disperse distributions while preserving 
the dominant particle size. 
We performed calculations over: 
51 values of the effective radius $r_{\rm{eff}}$
from 0.001 to 100 $\mu$m; 5 values of the real part of the
refractive index, $n_{\rm{r}}$=1.1, 1.5, 2, 2.5 and 3; 
14 values of the imaginary part of the refractive index, 
$n_{\rm{i}}$=0, 0.00001, 0.00003, 0.00006,    
             0.0001,  0.0003,  0.0006,  
             0.001,  0.003,  0.006,  
             0.01,  0.03,  0.06
             and 0.1. 
We also included the combinations 
$n_{\rm{r}}$=1.6 and $n_{\rm{i}}$=10$^{-4}$
(relevant to Mg$_2$SiO$_4$ and MgSiO$_3$), 
$n_{\rm{r}}$=2.25 and $n_{\rm{i}}$=10$^{-4}$
(relevant to CaTiO$_3$), and
$n_{\rm{r}}$=1.5 and $n_{\rm{i}}$=10$^{-7}$
(relevant to SiO$_2$). 
All calculations assumed a wavelength of 0.65 $\mu$m appropriate to the 
passband central wavelength of \textit{Kepler}.
Figure S6 presents a selection of 
$\varpi_{\rm{0}}$($r_{\rm{eff}}, n_{\rm{r}}, n_{\rm{i}}$) and $g$($r_{\rm{eff}}, n_{\rm{r}}, n_{\rm{i}}$) curves.

   \begin{figure*}[h]
   \centering
   \includegraphics[width=15.cm]{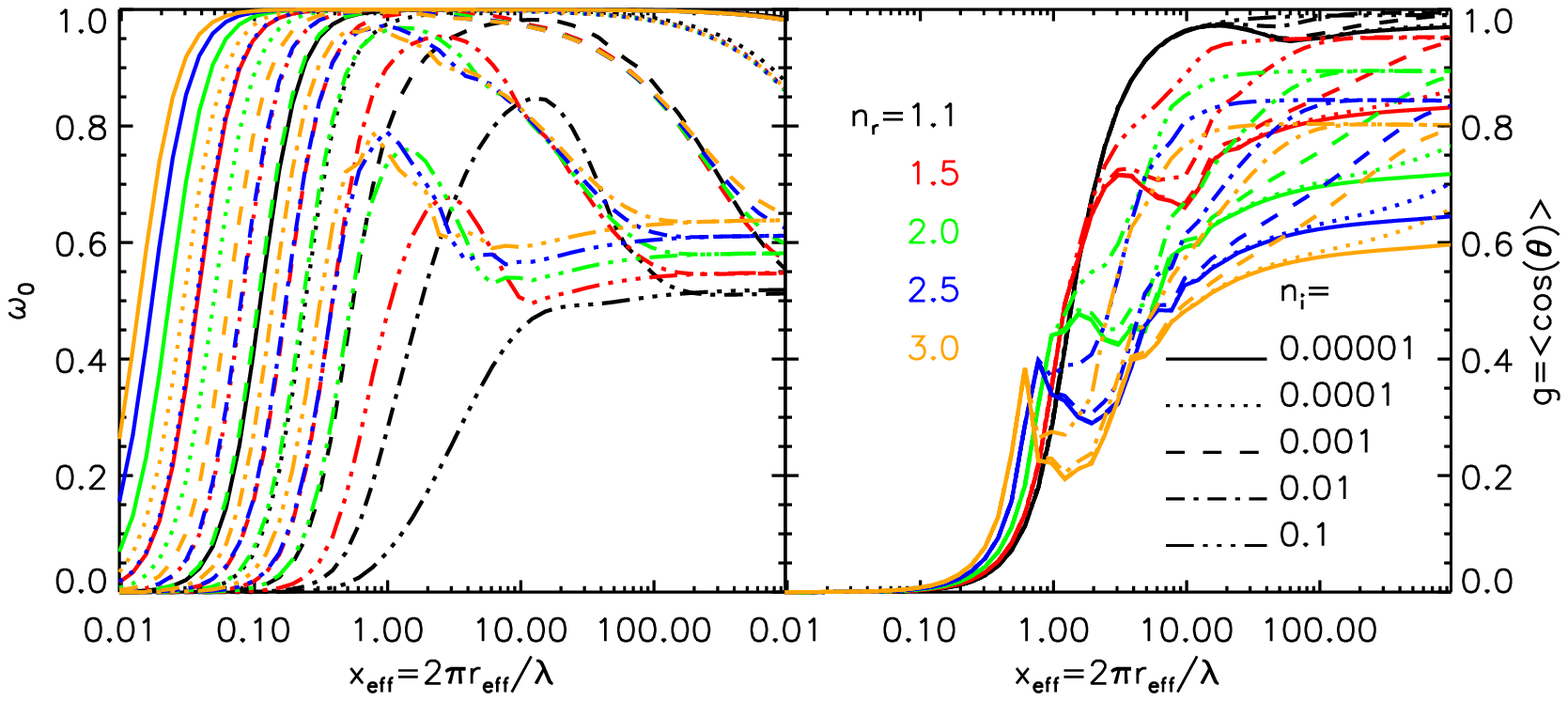}
   \caption*{\label{reffg_fig} Figure S6. 
   \textit{Mie theory calculations.} 
    Curves for $\varpi_{\rm{0}}$ and $g$ are produced and their dependences on 
    $r_{\rm{eff}}$, $n_{\rm{r}}$ and $n_{\rm{i}}$ investigated.
   }
   \end{figure*}


Figure 1 of Ref. \cite{wakefordsing2015} compiles refractive indices for exoplanet atmosphere condensates. 
We considered only substances that condense at temperatures of 1,000--2,000 K and whose optical properties can be determined in the optical, namely: 
SiO$_2$, Al$_2$O$_3$, FeO, CaTiO$_3$, Fe$_2$O$_3$, 
Mg$_2$SiO$_4$ (both Fe-rich and -poor), 
MgSiO$_3$, Na$_2$S, MnS, and TiO$_2$. Of these, only 
SiO$_2$, CaTiO$_3$, Fe-poor Mg$_2$SiO$_4$ and MgSiO$_3$ meet the condition 
$n_{\rm{i}}$$\lesssim$0.003 that we inferred from the model input parameters $\varpi_{\rm{0}}$ and $g$
within the minimal-$\chi^2$ sets. 
It is therefore justified to rule out the other potential condensates at the 99.99\% confidence level.
\\


\textbf{$A_{\rm{s}}$ vs. $A_{\rm{g}}$ diagram}

The geometric and spherical albedos are often-used measures of the overall planet
reflectance. 
The $A_{\rm{g}}$--$A_{\rm{s}}$ relationship for uniform atmospheres
has been investigated in past work \cite{hovenierhage1989, dlugachyanovitskij1974, vandehulst1974}. 
Our grid of model phase curves explores this relation
in a broader range of configurations that includes non-uniform cloud coverage. 
The geometric albedo $A_{\rm{g}}$ is defined as the planet net reflectance for a zero phase angle relative to the 
net reflectance of a Lambert disk with the same projected area. 
The spherical albedo is, by construction, $A_{\rm{s}}$$\le$1. 
Figure S7 provides a graphic summary of the calculated albedos, 
organised by colours according to the 
asymmetry parameter $g_1$ of the forward component. 
In our DHG parameterisation, 
increasing $g_1$ values lead to increased backscattering, which typically
results in enhanced $A_{\rm{g}}$ values. 

   \begin{figure}[h]
   \centering
   \includegraphics[width=12.cm]{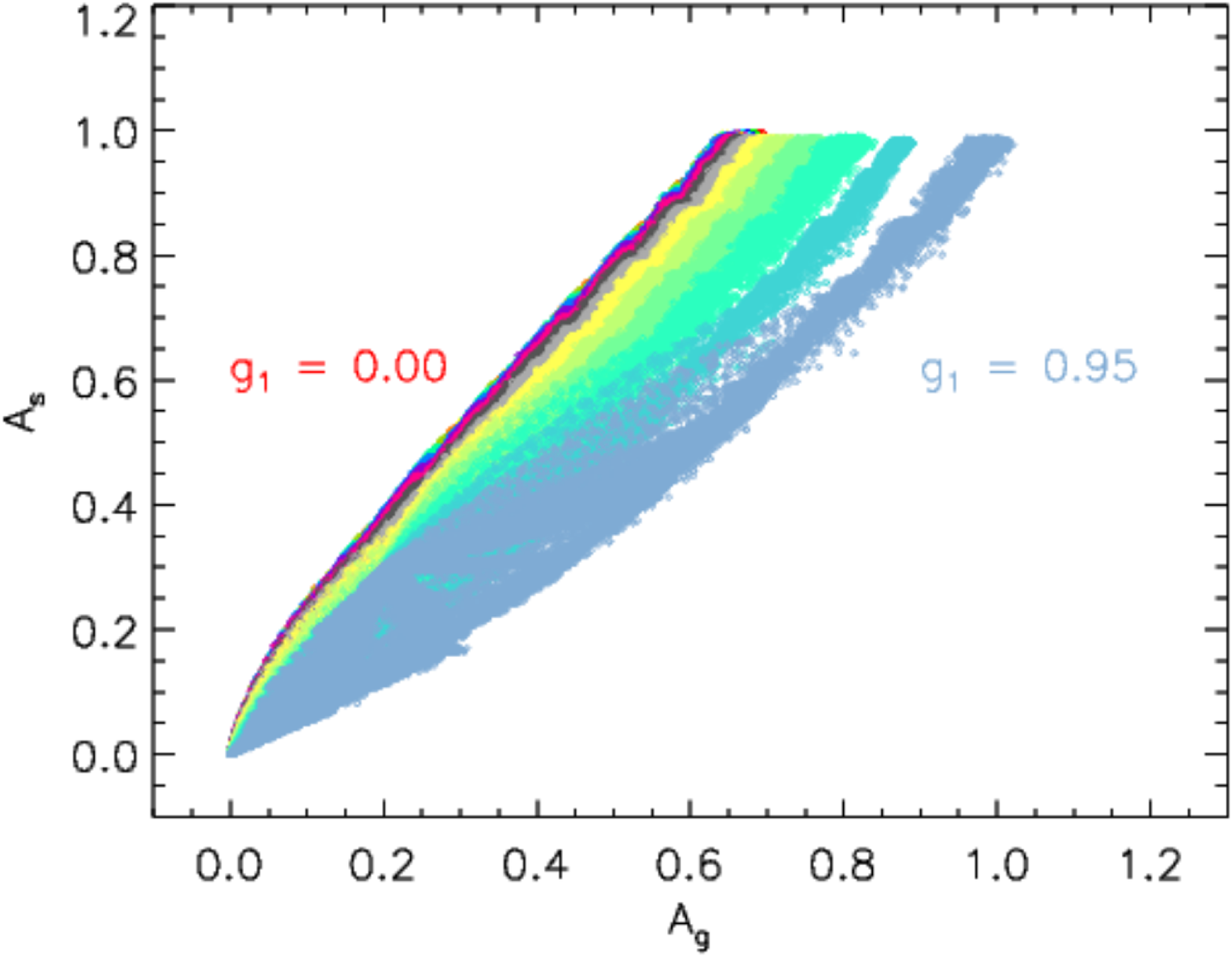}
   \caption*{\label{agvsas_fig} Figure S7. 
   \textit{Geometric and spherical albedos for model phase curves 
   with $r_{\rm{g}}$=0 and 0.3. DHG parameterisation.}
   }
   \end{figure}


   \begin{figure*}[t]
   \centering
   \includegraphics[width=16cm]{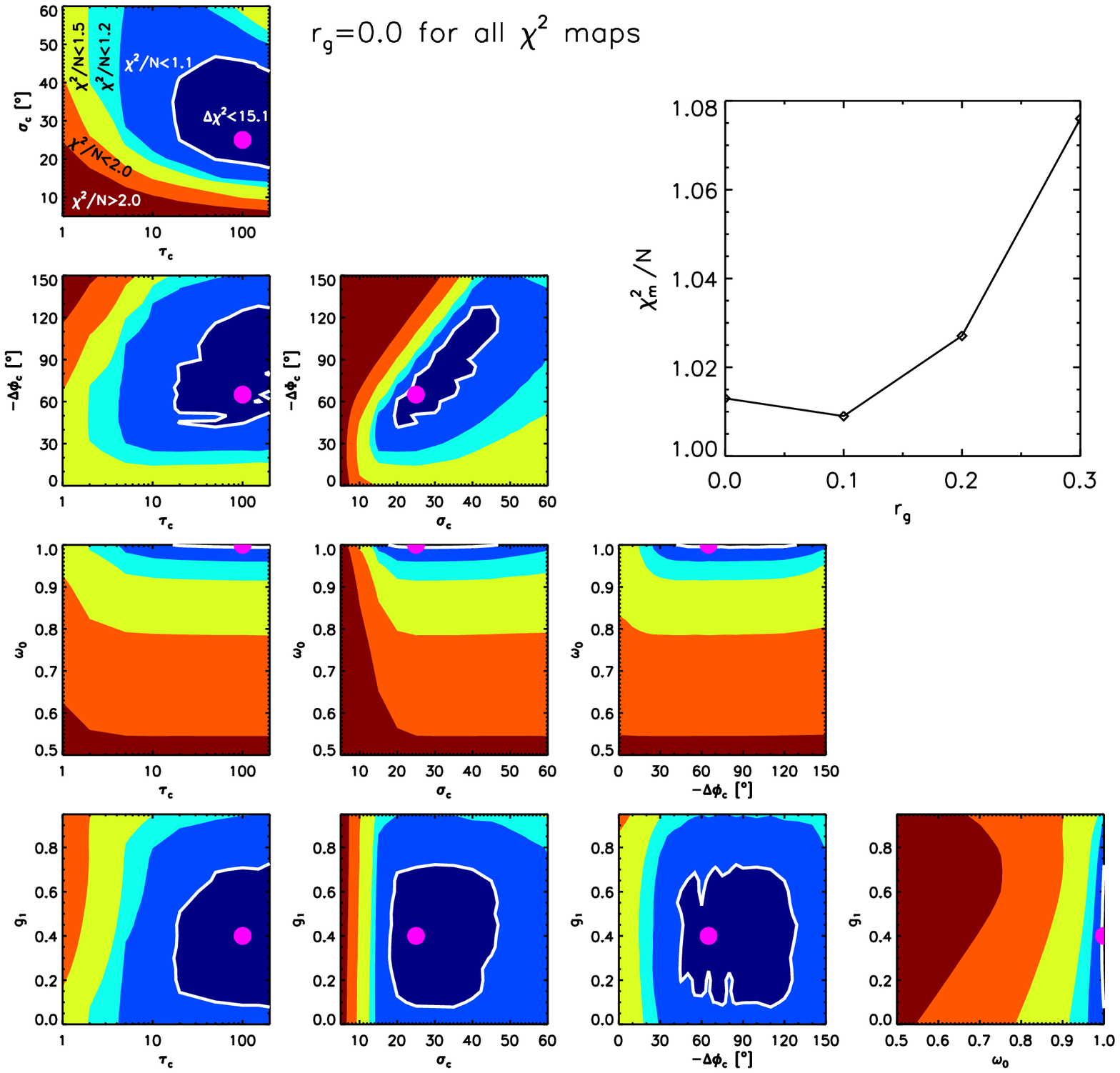}
   \caption*{Figure S8. 
   \textit{ Same as Fig. 2, but for a surface reflectance $r_{\rm{g}}$=0. }
   }
   \end{figure*}

   \begin{figure*}[t]
   \centering
   \includegraphics[width=16cm]{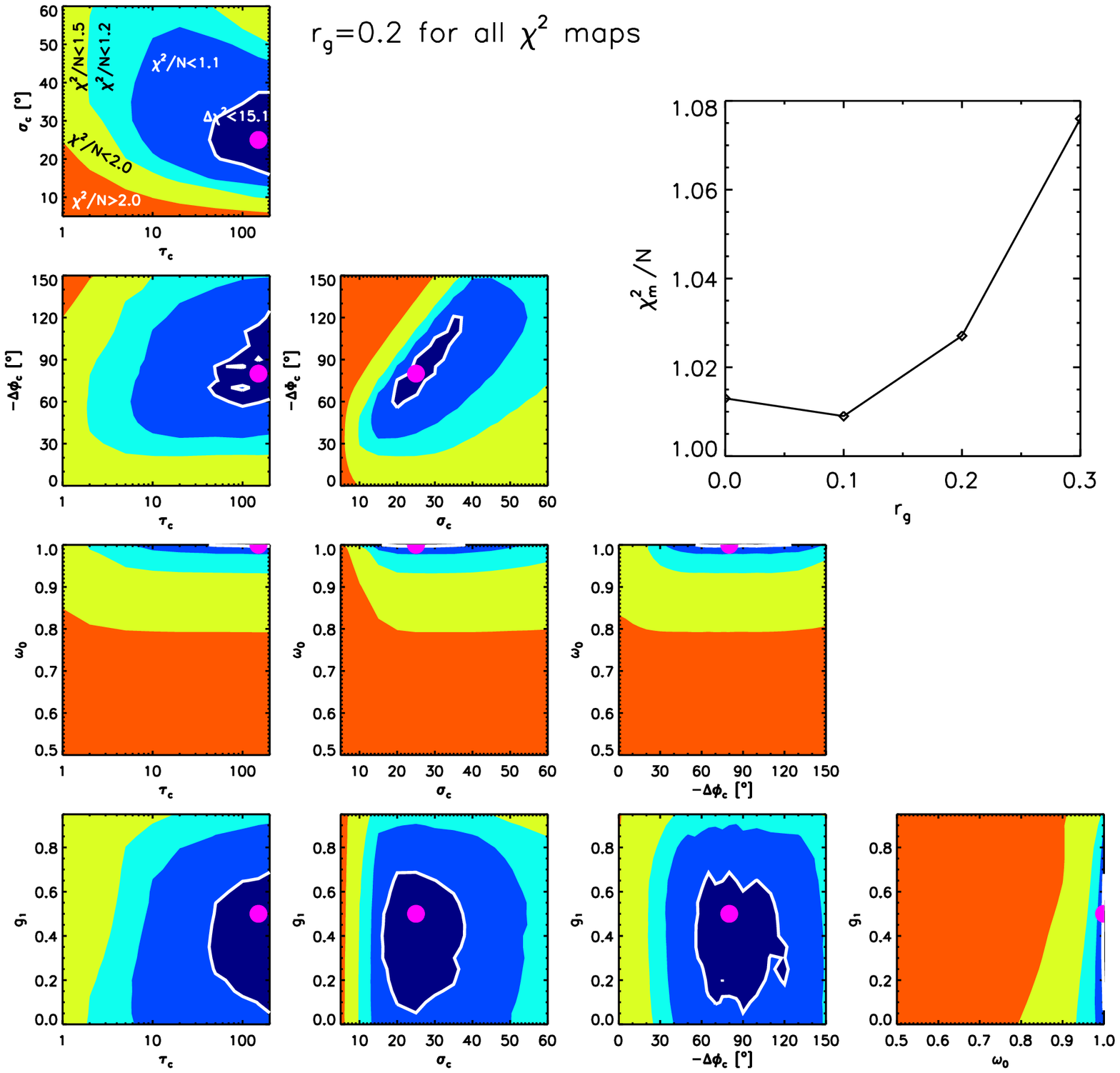}
   \caption*{Figure S9. 
   \textit{ Same as Fig. 2, but for a surface reflectance $r_{\rm{g}}$=0.2. }
   }
   \end{figure*}

   \begin{figure*}[t]
   \centering
   \includegraphics[width=16cm]{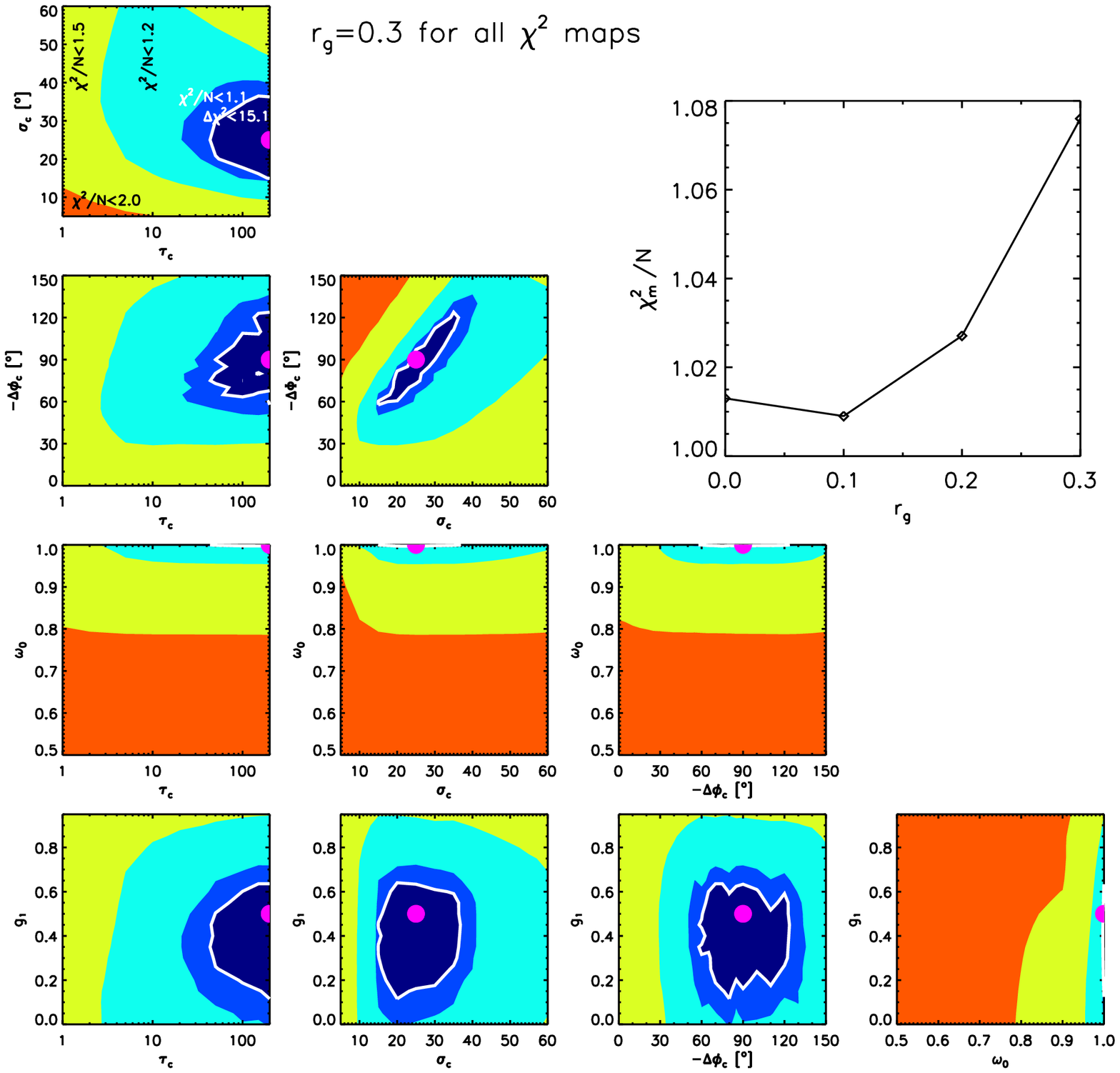}
   \caption*{Figure S10. 
   \textit{ Same as Fig. 2, but for a surface reflectance $r_{\rm{g}}$=0.3. }
   }
   \end{figure*}

\cleardoublepage